\documentclass[a4paper,10pt]{article}

\usepackage{amssymb}
\usepackage{amsfonts}
\usepackage{amsmath}
\usepackage{subfigure}
\usepackage{algorithm}
\usepackage[noend]{algorithmic}
\usepackage{amsmath,amssymb,cite}
\usepackage{epsfig}
\usepackage{pst-node,pstricks}

\usepackage{graphicx}
\usepackage{psfrag}
\usepackage[nohead, top=2.9cm, bottom=3.2cm, left=2.9cm, right=3.1cm]{geometry}

\newtheorem{definition}{Definition}[section]

\def\q5uad{\quad\quad\quad\quad\quad}

\title{Sanitize and Response-time Bounds for Malware \\ Pandemics in Mobile Devices \vspace{0.2cm}}

\title{Counter-measure Response Time-Bounds for Malware Pandemics Prevention in Mobile Devices \vspace{0.2cm}}

\title{Preventing Malware Pandemics in Mobile Devices \\ by Establishing Response-time Bounds  \vspace{0.2cm}}

\author{Stavros~D.~Nikolopoulos \ \ Iosif~Polenakis}

\date{}

\begin{document}

\maketitle

\vspace{-0.5cm}

\centerline{\it Department of Computer Science \& Engineering}

\centerline{\it University of Ioannina}

\centerline{\it GR-45110 \ Ioannina, Greece}

\centerline{\tt \{stavros,ipolenak\}@cs.uoi.gr}


\vskip 0.3in

\begin{abstract}
\noindent We study the propagation of a malicious software in a network of mobile devices, which are moving
in a specific city area, and establish time bounds for the activation of a counter-measure, i.e.,
an antivirus or a cleaner in order to prevent pandemic. More precisely, given an initial
infected population (mobile devices), we establish upper bounds on the time needed for a counter-measure
to take effect after infection (response-time), in order to prevent the rest susceptible devices
to get infected. Thus, within a period of time, we guarantee that not all the susceptible devices
in the city get infected and the infected ones get sanitized. In our work, we first
propose a malware propagation model along with a device mobility model and then, utilizing these
models, we develop a simulator that we use to study the spread of malware in such networks.
Finally, we provide experimental results for the pandemic prevention taken by our simulator
for various response-time intervals.

\vspace*{0.1in}
\end{abstract}

\vspace*{0.1in}
\section {Introduction}
\label{sec:Introduction}
\vspace*{0.05in}

\noindent A malicious software or  \emph{malware} may refer to any kind of software that its functionality causes harm to a user, computer, or network. The motivation of our research is triggered by the enormous grow and spread on the number of malicious software \cite{MaAvKo}, and much more, on mobile devices. The main difference between networks formed by devices connected via ethernet and networks formed by mobile devices, is that the former are static while the later ones are dynamic networks (i.e. networks that their representing graph changes during time - {\it ad-hoc} networks).

\vspace*{0.1in}
\noindent \textbf{Epidemic Models} Epidemic models can be applied to any network structures to describe the propagation of a disease despite of its type (i.e., biological virus or computer virus) between a set of entities. The overall propagation can be described as a branching process, e.g., a tree that its root is the initial infected population and every level contains child nodes representing the population infected by the nodes of the previous level.

Briefly speaking, such epidemic models describe the nodes - entities by a set of potential states or conditions they can go through the course of the epidemic, namely {\tt Susceptible, Infected, Repaired, Removed, Immune}. In the {\tt Susceptible} state a node is potentially vulnerable to a disease, while when the node gets infected (probably by its neighbors) then it goes to {\tt Infected} state. On the other hand, depending on the modeled cases, if the disease is destructive for its host then after a period of time the infected node goes to {\tt Removed} state, while if a cure exist and is been applied to an infected node then after a period of time (throughout this paper we shall call it sanitize-time) the node goes to {\tt Repaired} state, where, depending again on the modeling demands, it can be either {\tt Immune} or not. Next, we briefly present various epidemic models that can be deployed according to the needs of the simulated problem.

\begin{itemize}
\item {\bf SI Epidemic Model} It the most trivial model containing only two states ({\tt Susceptible, Infected}). Once a node is susceptible and gets infected (and hence infectious), then it remains forever in this state. The following epidemic models considered as variations of the SI model. \cite{AvMaCh,AvMaSt,DrRo,EaKl}.

\item {\bf SIRp Epidemic Model} In this model an infected node can be repaired in some fashion \cite{AvMaCh,AvMaSt,EaKl}. To this point, it is worth noting that a node repair may provide immunization to the host against the disease or not. Depending on this fact, the following specifications arise as special cases of SIRp model:
        \begin{itemize}
        \item[$\circ$] {\bf SII Epidemic Model} The SII model (last I stands for Immune) results as a solution when we need to formally describe the propagation of a disease where there exists a cure that immunizes/sanitizes \cite{YaChe} the infected hosts after their treat \cite{EaKl,GaGo,SaKa}.
        \item[$\circ$] {\bf SIS Epidemic Model} On the other hand, the SIS model (last S stands for Susceptible) is suitable for the cases where even though exists a cure for the infected node, it still stays susceptible on getting the disease again \cite{EaKl}
         \item[$\circ$] {\bf SIRS Epidemic Model} The SIS model is also referred as SIRS where the `R' stands for {\tt Repaired} (i.e. `Rp' in our case). A further specification may be appeared extending SIS model depending on the case and the demands of the situation under modeling.
        \end{itemize}

\item {\bf SIRm Epidemic Model} Finally, if the modeled disease is destructive for the infected host, i.e., no cure exists, then SIRm model (last Rm stands for Removed) is suitable for application in such case to model the epidemic.

\item {\bf SEIS Epidemic Model} The SEIS model introduces a new state (i.e. {\tt Exposed})takeing into account the latent period of a disease where a node may be exposed to the disease by, for example, a close contact with an infected node. In this model, any immunity has been left to an infected node leaving an infected node to be susceptible again in the time after the infection.

\item {\bf SEIR Epidemic Model} However, similarly to the case of SIRp model, the SEIR epidemic model formally describes the propagation of a disease where there exists a cure that immunizes or simply repairs an infected host, once has been firstly exposed and then infected by the disease.
\end{itemize}

\vspace*{0.1in}
\noindent \textbf{Related Work} \ In \cite{BoShi}, Bose and Shin investigate the propagation of mobile worms and viruses that spread primarily via SMS/MMS messages and short-range radio interfaces such as Bluetooth. In this work, they study the propagation of a mobile virus similar to Commwarrior in a cellular network using data from a real-life SMS customer network, modeling each handheld device as an autonomous mobile agent capable of sending SMS messages to others (via an SMS center) and capable of discovering other Bluetooth equipped devices. Their results show that hybrid worms that use SMS/MMS and proximity scanning (via Bluetooth) can spread rapidly within a cellular network.

An interaction-based simulation framework to study the dynamics of worm propagation over wireless networks developed by Channakeshava et al. \cite{ChaCha}. This framework is constructed by their proposed methods for generating synthetic wireless networks using activity-based models of urban population mobility. With this framework they study how Bluetooth worms spread over realistic wireless networks.

In \cite{ChJi}, Chen and Ji focus on modeling the spread of topological malware (spreads based on topology information), as to understanding its potential damages, and developing countermeasures to protect the network infrastructure.
Their model is motivated by probabilistic graphs, using a graphical representation to abstract the propagation of malwares that employ different scanning methods. Utilizing a spatial-temporal random process they describe the statistical dependence of malware propagation in arbitrary topologies. Finally, their results show that the independent model outperforms the previous models, whereas the Markov model achieves a greater accuracy in characterizing both transient and equilibrium behaviors of malware propagation.

Fleizach et al. \cite{FlLiJo}, evaluate the effects of malware propagating using communication services in mobile phone networks. Although self-propagating malware is well understood in the Internet, mobile phone networks have very different characteristics in terms of topologies, services, provisioning and capacity, devices, and communication patterns. To investigate malware in mobile phone networks, they developed an event-driver simulator that captures the characteristics and constraints of mobile phone networks, modeling realistic topologies and provisioned capacities of the network infrastructure, as well as the contact graphs determined by cell phone address books.

\vspace*{0.1in}
\noindent \textbf{Our Contribution} \ In this paper we investigate the effect caused by the response-time of a counter-measure (i.e., an antivirus or a cleaner) on the propagation of a malware in mobile devices. For this purpose, we first propose a malware propagation model along with a device mobility model and then we develop a simulator that we use to study the spread of malware in the network formed by the mobile devices \cite{NiPo2}. Our malware propagation model is based on the SIRp epidemic model where the devices can be either in {\tt Susceptible}, {\tt Infected} or {\tt Repaired} ({\tt Immunized}) state. Additionally, our device mobility model generates traces, utilizing shortest path algorithms, for the mobile devices which are moving inside a city. Note that the city is represented by its image taken from Google Maps and its town-planning is modeled by an $n \times m$  matrix of $0s$ and $255s$, where $0$ denotes a point on a road while $255$ denotes any obstacle. We utilize our malware propagation and device mobility models and develop a simulator that we use to study the spread of malware in mobile devices.

Given an initially infected population, we perform a series of simulations for various response-time intervals, taking into account other factors which affect the malware's propagation, i.e., initially infected population and network density, and establish upper bounds on the response-time needed by a counter-measure, such as a malware detector \cite{NiPo1}, to take effect as to prevent pandemic. In other words, through our model we determine the maximum permitted time for a counter-measure to be activated when a specific percentage of the population is infected in order to guarantee that not all the susceptible devices in the city get infected and some (or, all) infected ones get sanitized. We finally present experimental results for the pandemic prevention provided by our simulations for various response-time intervals.

\vspace*{0.1in}
\noindent \textbf{Road Map} The paper is organized as follows. In  Section~2 we present a malware propagation model to simulate the spread of malware based on geological proximity and, then, a device mobility model that generates traces for the mobile devices, utilizing shortest path algorithms, inside the city represented by its image taken from Google Maps. In Section~3 we proceed by evaluating our model and present results achieved by our simulator that deploys this model.
Finally, Section~4 concludes the paper and discusses issues for farther investigation.

\vspace*{0.1in}
\section {Model Design}
\label{sec:model_design}
\vspace*{0.05in}

\noindent In this section, we present our propagation model simulating the spread of a malware to proximal mobile devices, as well as the mobility model and its main principles concerning the motion of the devices in a city. Additionally, we show the representation of town-planning through an $n \times m$ matrix consisted by $0s$ and $255s$, where $0$ denotes a point on a road of the city and $255$ denotes any obstacle (e.g., buildings) produced by an image taken from Google Maps.

\vspace*{0.1in}
\subsection{City Representation}
\begin{figure}[t!]
\hrule\medskip\medskip
  \begin{minipage}[b]{0.5\linewidth}
    \centering
    \includegraphics[width=.8\linewidth]{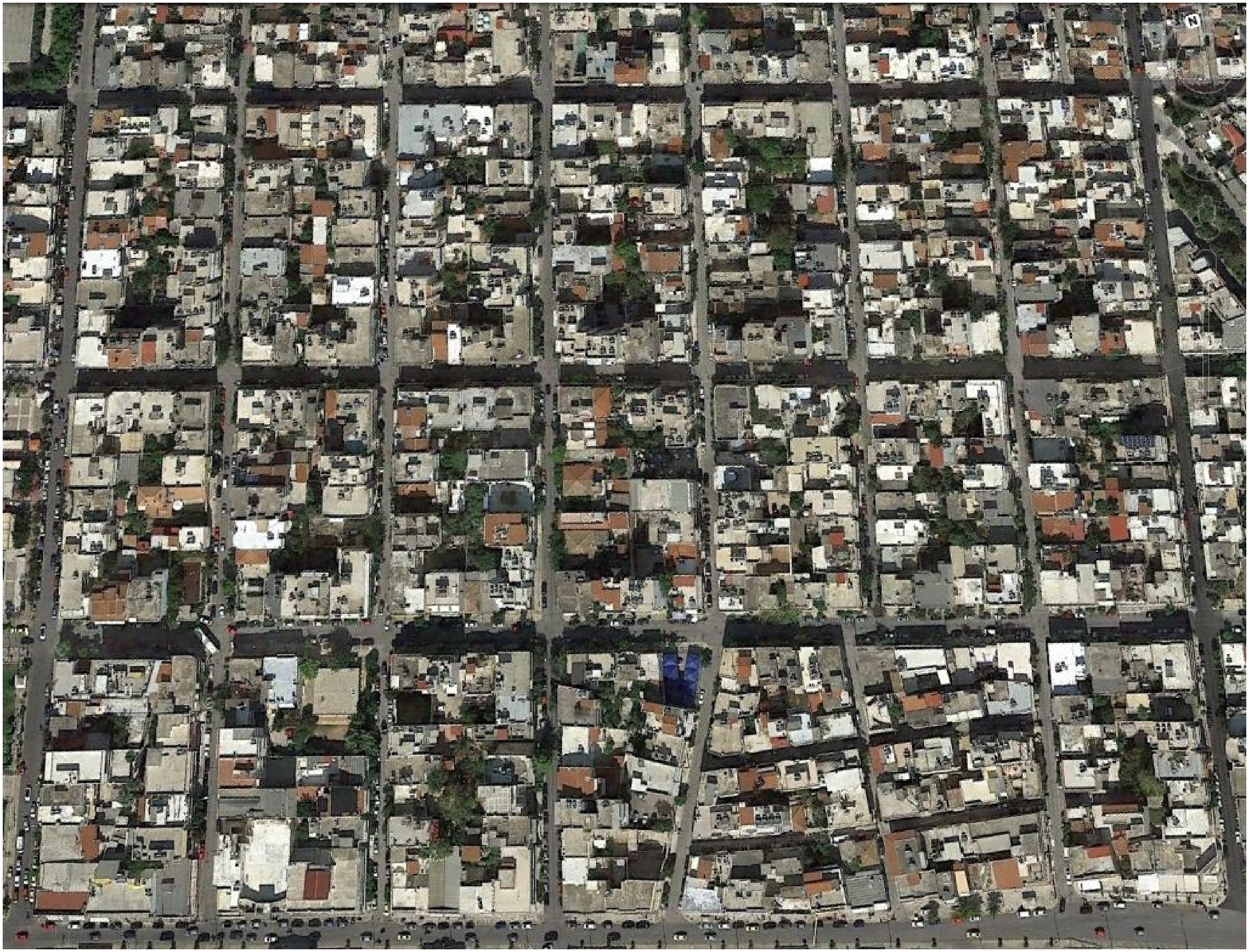}\\
     \smallskip {\small $(a)$ Initial image from Google Maps}
    \vspace{2ex}
  \end{minipage}
  \begin{minipage}[b]{0.5\linewidth}
    \centering
    \includegraphics[width=.8\linewidth]{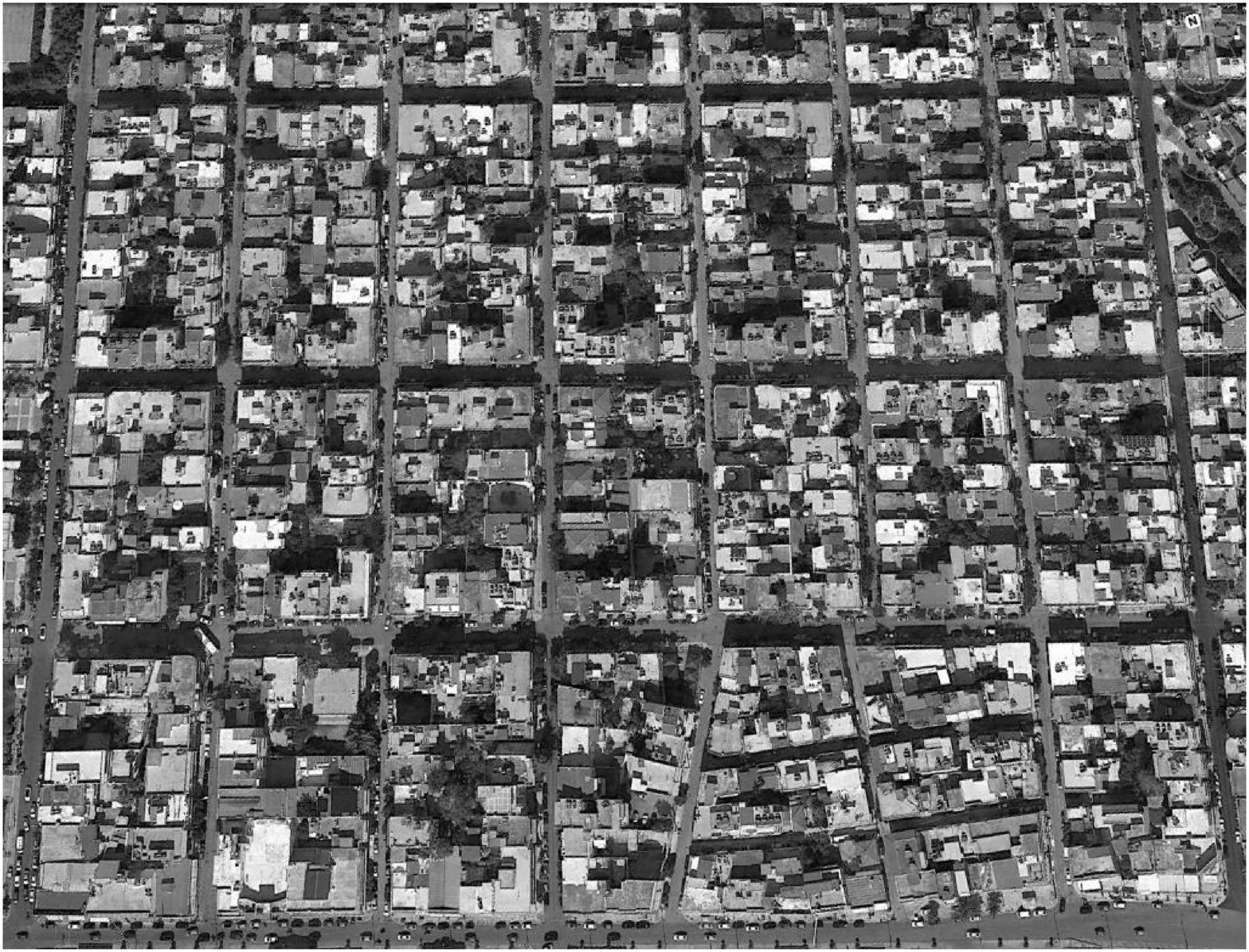}\\
     \smallskip {\small $(b)$ Gray-scale image}
    \vspace{2ex}
  \end{minipage}
  \begin{minipage}[b]{0.5\linewidth}
    \centering
    \includegraphics[width=.8\linewidth]{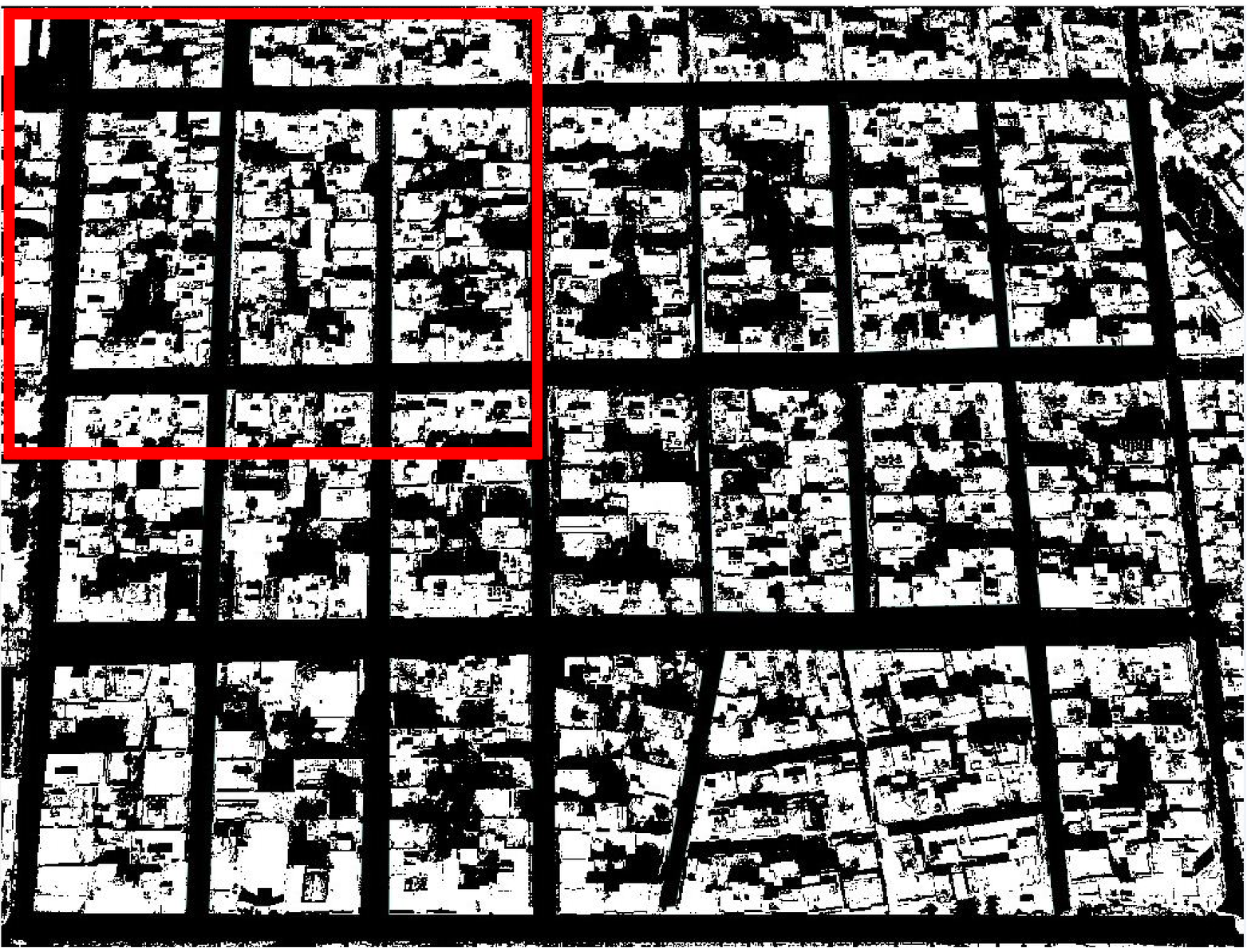}\\
    \smallskip {\small $(c)$ Black and White image}
    \vspace{2ex}
  \end{minipage}
  \begin{minipage}[b]{0.5\linewidth}
    \centering
    \includegraphics[width=6.4cm,height=4.6cm,]{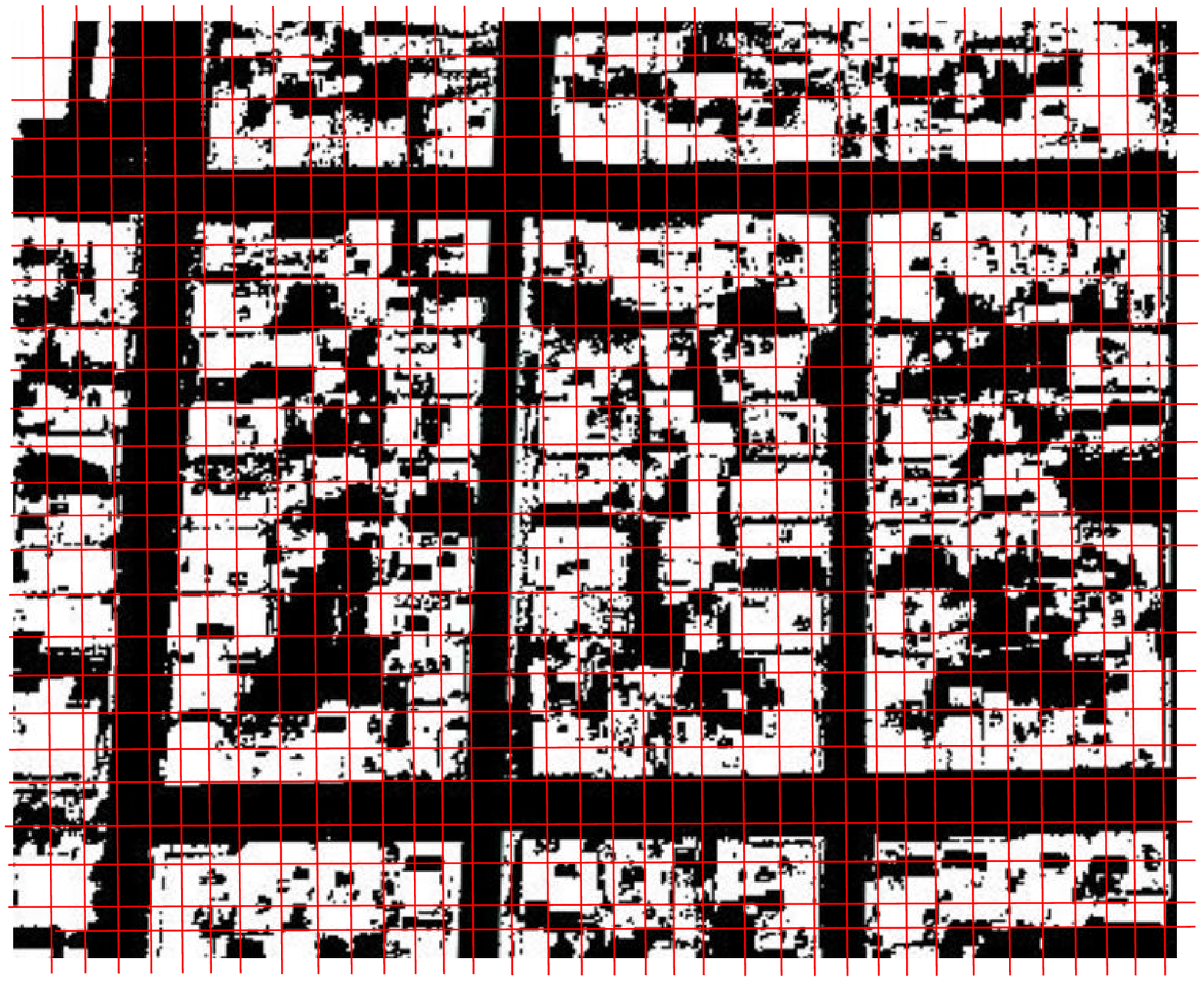}\\
     \smallskip {\small $(d)$ Grid-snapshot of the Black and White image}
    \vspace{2ex}
  \end{minipage}
  \hrule\smallskip
  \caption{\small{$(a)-(c)$ The transformation of a Google Maps image to a Black and White matrix; $(d)$ Zoom-in to a part of a Black and White image.}}
  \label{fig:city}
\end{figure}

Our model simulates malware propagation to mobile devices that are changing their positions, or geological coordinates, according to a town-planing. In order to make our simulation closer to reality we used images of real towns-planning from Google Maps. We transform these images from RGB to gray-scale color system and then to black and white using an appropriate threshold. Hence, having an image of dimensions $n \times m$ representing the town-planing of the city, we transform it to an $n \times m$ matrix $M_{map}$ with values $0s$  and $255s$, where $0$ represent a free space (i.e., road) and $255$ represents any obstacle (i.e. building). So, in our simulation, we permit a mobile device to move into a position with coordinates $(x,y)$ if the corresponding cell $(i,j)$ of matrix $M_{map}$ has value $0$. Additionally, we have assigned a weight to each cell with value $0$ in $M_{map}$ matrix to represent its level of attraction, being hence a {\tt cold-}, {\tt warm-} or {\tt hot-} {\tt spot}, with weights $w=1$, $w=5$, and $w=10$ respectively. Next, we will discuss how these values are used in order to compute a path from one point to another using a shortest path algorithm. In Figure~\ref{fig:city} we illustrate the construction of the town-planning representation. 

\vspace*{0.1in}
\subsection{Malware Propagation Model}
The proposed malware propagation model for simulating the malware's spread to neighboring mobile devices has its basis to factors that effect the spread such as the {\tt range} of a mobile device, the {\tt size} of a malware in terms of packets, and the {\tt velocity} of a node. In order to get infected, a mobile device should have collect all the packets that constitute the malware. In our model for simulating the propagation of malware we allow the activation of a counter-measure to remove the malicious software from the infected devices. This feature is adapted to our model in order to remove, or clean in some fashion, the malicious software from the device. Specifically, in our model, if a device remains for a period of time (i.e., simulation steps demanded for the transmission of all packets of the malware) within a specific radius from an infected device then it gets infected too. However, if this device moves out of range then it would need more time, in terms of simulation steps, in order to get infected. Upon the activation of a counter-measure, it sanitizes the device by removing the malware.

In our model for simulating malware propagation in mobile devices, it is crucial to take into account that such networks are formed {\tt on-the-fly} between the devices while they are moving inside the city. This network can be represented by an undirected graph (we shall denote it $G_{dev}$ throughout the paper) that is modifying its structure (i.e. topology by means of edge creations and deletion among its nodes - devices). So, we could claim that during a specific period of time, let $[t_1,t_n]$ this graph can be referenced by its structurally different instances as ${G_{dev}^1,G_{dev}^2, ..., G_{dev}^n}$.

\begin{definition}{}
We define $G_{dev}=(I,S,E)$ to be a bipartite graph whose vertices correspond to the devices of the network and an edge between two vertices occur if their corresponding devices have distance less than $r$ at time $t$.
\end{definition}

\begin{figure}[t!]
\hrule\medskip\medskip
  \begin{minipage}[b]{0.5\linewidth}
    \centering
    \includegraphics[width=.9\linewidth]{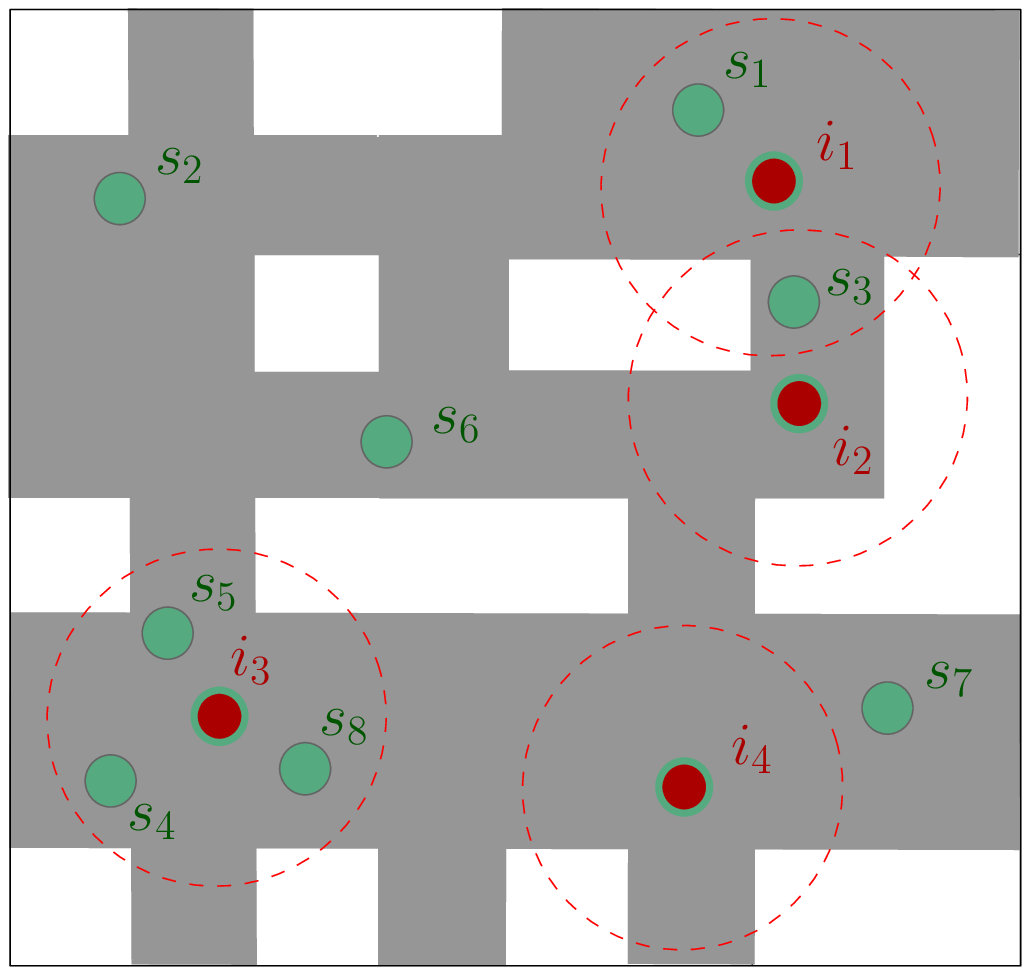}\\
     \smallskip {\small $(a)$ Probing devices}
    \vspace{2ex}
  \end{minipage}
  \begin{minipage}[b]{0.5\linewidth}
    \centering
    \includegraphics[width=.9\linewidth]{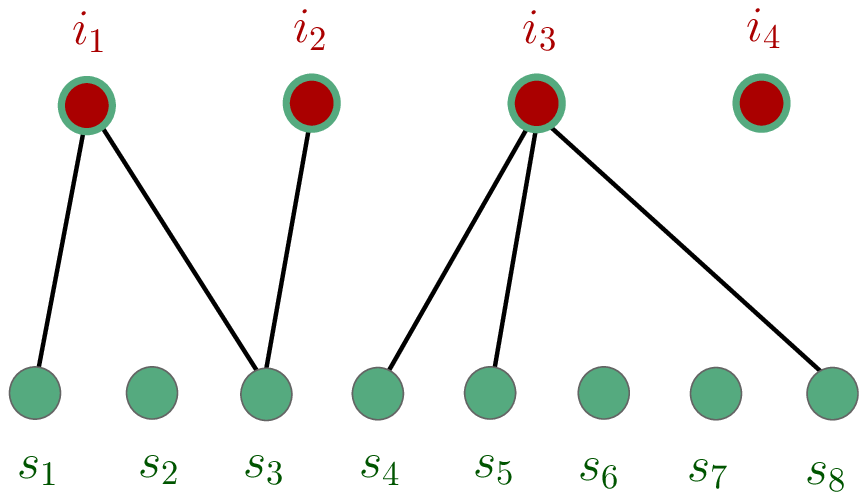}\\
    \vspace{0.5in}
     \smallskip {\small $(b)$ The graph $G_{dev}$}
    \vspace{2ex}
  \end{minipage}
  \hrule\smallskip
  \caption{\small{Mobile devices moving inside a city represented by its $M_{map}$ matrix, constructing the bipartite graph $G_{dev}$ by the links formed among them.}}
  \label{fig:gdev}
\end{figure}

\noindent In Figure~\ref{fig:gdev} we illustrate the process of constructing the bipartite graph $G_{dev}$. In Figure~\ref{fig:gdev}($a$) we depict how the mobile devices are moving inside a city represented by its $M_{map}$ matrix; recall that cells  with value $0$ correspond to roads while cells with value $255$ correspond to obstacles. The circles around the mobile devices show the range of them while their colors, blue or red, correspond to transmissions by susceptible or infected devices respectively. Then, in Figure~\ref{fig:gdev}($b$), we illustrate how we create the bipartite graph $G_{dev}$: for any pair of mobile devices (i.e., vertices in the $G_{dev}$), we add an edge between them in $G_{dev}$ if their distance is less than $r$. Concerning the above network represented by $G_{dev}$, next we provide some definitions about its characteristics.

\begin{definition}{}
Let $\tau_1, \tau_2, ..., \tau_k$ be the $k$ states of a given epidemic model $E$. We define the {\it state-cover} $C(\tau^*)$ of state $\tau^* \in \{\tau_1, \tau_2, ..., \tau_k\}$ as the rate of the objects (i.e., mobile devices) that are in state $\tau^*$ by the sum of the objects that are in any state of the epidemic model $E$, that is

\begin{equation}\label{density}
    C(\tau^*)=\dfrac{|\tau^*|}{\sum_{i=1}^{k}|\tau_i|}.
\end{equation}

\end{definition}

\vspace{0.1in}
\noindent In our model, for the {\tt infected} state $I$ and the {\tt susceptible} state $S$, we compute the {\it Infected-cover} and the {\it Susceptible-cover} as follows:

\begin{equation}\label{density}
    C(I)=\dfrac{|I|}{|S|+|I|+|Rp|} \ \ \ \ {\text {and}} \ \ \ \   C(S)=\dfrac{|S|}{|S|+|I|+|Rp|},
\end{equation}

\vspace{0.1in}

\noindent respectively, where $|Rp|$ is the number of devices in the {\tt repaired} state $Rp$ of our model.

\begin{definition}{}
Let $\tau_1^*, \tau_2^* \in \{\tau_1, \tau_2, ..., \tau_k\}$ be two states of a given epidemic model $E$. We define the {\it state-rate} $R(\tau_1^*\tau_2^*)$ of states $\tau_1^*$ and $\tau_2^*$ as the rate of the number of the objects belonging to $\tau_1^*$ state over the number of objects that belong to $\tau_1^*$ state plus 1, that is

\begin{equation}\label{density}
   R(\tau_1^*, \tau_2^*)=\dfrac{|\tau_1^*|}{|\tau_2^*|+1}.
\end{equation}

\end{definition}

\vspace{0.1in}
\noindent In our model, the state-rates {\it IS-rate} and {\it SI-rate} for the {\tt infected} state $I$ and the {\tt susceptible} state $S$ are the following:
\begin{equation}\label{density}
    R(I,S)=\dfrac{|I|}{|S|+1} \ \ \ \ {\text {and}} \ \ \ \   R(S,I)=\dfrac{|S|}{|I|+1},
\end{equation}

\vspace{0.1in}

\noindent respectively.

\vspace*{0.1in}
\subsection{Device Mobility Model}
In our model, we simulate the movements of a mobile device by changing the coordinates of a node taking into account the corresponding cells $(i,j)$ in the $M_{map}$ matrix with values $0$ and $255$ that represents the town-planning. More precisely, we permit a device to move on a point in the map if the corresponding cell $(i,j)$ in $M_{map}$ matrix has value $0$ since such a cell represents a road or, equivalently, we do not allow a device to move on a cell with value $255$ since it represents any obstacle such as building. To make our simulation more realistic, we propose and implement a trace generator for device mobility, that is, for each node we generate a trace between an initial position and a target position. In particular, we set each node at a pre-defined point with coordinates $(i,j)$ on the grid and then a destination point $(i',j')$ is assigned on that as to be reached through a path computed by a shortest path algorithm computed on the weighted directed graph we define next, and which we shall call it $G_{map}$.

\begin{definition}{}
The weighted directed graph $G_{map}$ represents the town's planning using its $M_{map}$ representation; it is constructed as follows:
\begin{itemize}
        \item[$\circ$] its vertices $V(G_{map})$ correspond to the $0$-value cells of matrix $M_{map}$, and
        \item[$\circ$] two vertices of $V(G_{map})$ are joined by an edge if their corresponding cells with value $0$ are adjacent in the $M_{map}$ matrix; note that, a $0$-value cell $(i,j)$ of $M_{map}$ is adjacent to every $0$-value cell in its 8-neighborhood. The weight of an edge $(u_i, v_i)$ in $E(G_{map})$ has value $10-w_{u_i} + 10-w_{v_i} + 1$, where $w_{u_i}$ and $w_{v_i}$ are the attraction levels of nodes $u_i$ and $v_i$, respectively.
\end{itemize}

\end{definition}

\begin{figure}[t!]
\hrule\medskip\medskip
\begin{minipage}[b]{0.5\linewidth}
  \begin{minipage}[b]{0.5\linewidth}
    \centering
    \includegraphics[width=.9\linewidth]{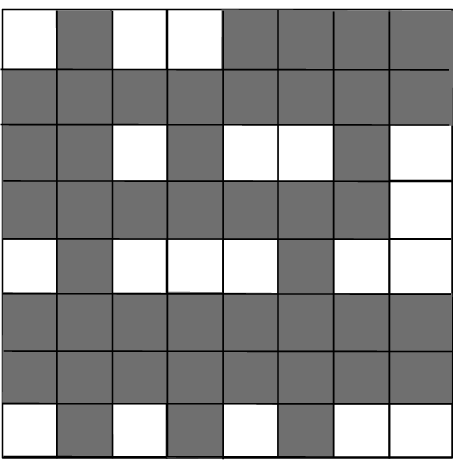}\\
     \smallskip {\small $(a)$ The matrix $M_{map}$}
    \vspace{2ex}
  \end{minipage}
  \begin{minipage}[b]{0.5\linewidth}
    \centering
    \includegraphics[width=.9\linewidth]{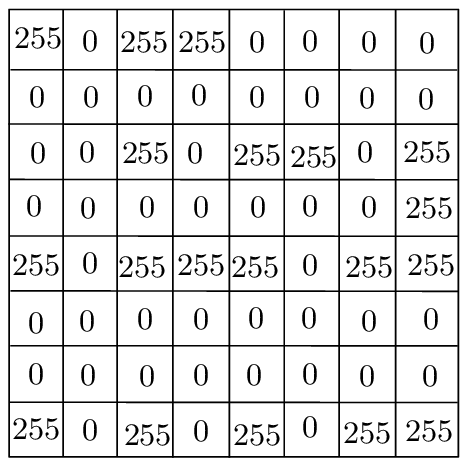}\\
     \smallskip {\small $(b)$ The $M_{map}$ in $[0,255]$}
    \vspace{2ex}
  \end{minipage}
  \begin{minipage}[b]{0.5\linewidth}
    \centering
    \includegraphics[width=.9\linewidth]{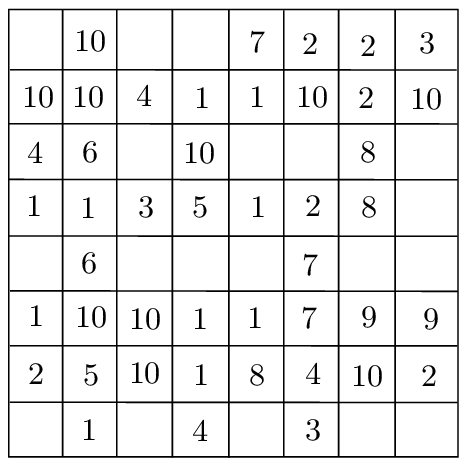}\\
    \smallskip {\small $(c)$ Attraction levels}
    \vspace{2ex}
  \end{minipage}
  \begin{minipage}[b]{0.5\linewidth}
    \centering
    \includegraphics[width=.9\linewidth]{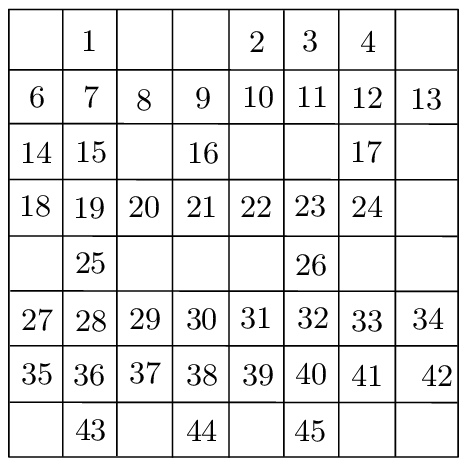}\\
     \smallskip {\small $(d)$ ID assignment}
    \vspace{2ex}
  \end{minipage}
\end{minipage}

\vspace{0.3in}

\centering
\begin{minipage}[b]{1\linewidth}
    \centering
    \includegraphics[width=.83\linewidth]{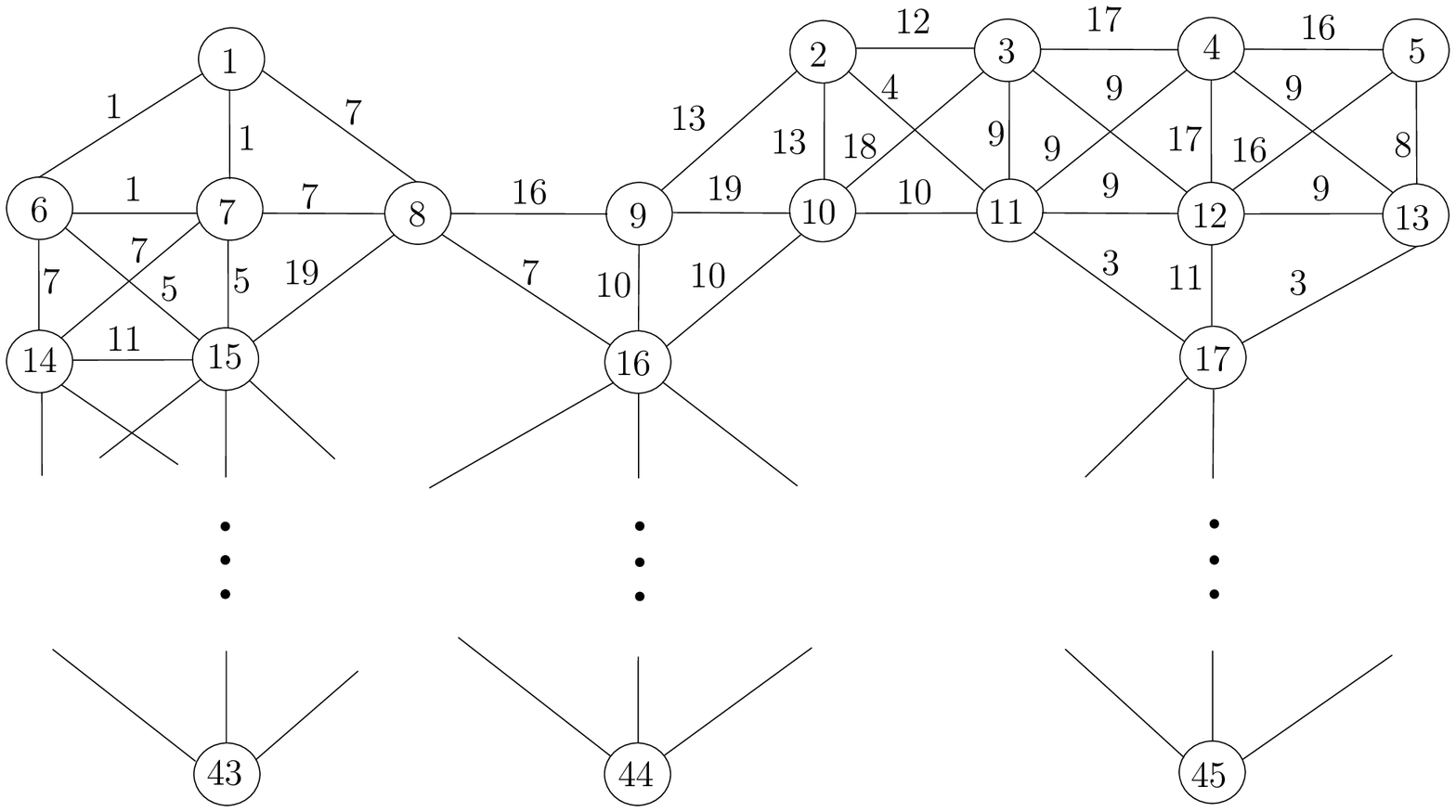}\\
     \medskip {\small $(e)$ The graph $G_{map}$}
    \vspace{2ex}
\end{minipage}

  \hrule\smallskip
  \caption{\small{Map representation and undirected weighted graph construction.}}
  \label{fig:gmap}
\end{figure}

\noindent In Figure~\ref{fig:gmap}, we show in detail the construction of the graph $G_{map}$ from the $M_map$ matrix. In particular, in Figures ~\ref{fig:gmap}($a$) and ~\ref{fig:gmap}($b$) we show the black/white representation of an example map representing each point of a road with value $0$ (black) and any obstacle (building) with value $255$ (white), in Figure~\ref{fig:gmap}($c$) we show the attraction level matrix which is constructed by assigning to its cells values in the range $[1,10]$ depicting the {\tt cold-}, {\tt warm-} and {\tt hot-} {\tt spot} of the city, in Figure~\ref{fig:gmap}($d$) we assign an ID on each point, indicating a node on the $G_{map}$ graph, while in Figure~\ref{fig:gmap}($e$) we present the resulting graph $G_{map}$.

To this point, it is worth noting to refer that the destination points are not randomly assigned as for each device we select a destination that is located to the {\tt NW}, {\tt N}, {\tt NE}, {\tt W}, {\tt E}, {\tt SW}, {\tt S}, {\tt SE} boundaries of the map, where once a destination reached by a mobile device, then a new destination point is assigned and thus we guarantee that always the devices change their positions.

Finally, deepen into the relation between the graphs $G_{map}$ and $G_{dev}$ we can claim that the structure of $G_{dev}$ strongly dependents on the structure of $G_{map}$. The relation between these two graphs relies on the property that the density of $G_{dev}$ (by means of sparse or dense graphs) is affected by the cardinalities of vertex sets $V(G_{dev})$ and $V(G_{map})$, as the cardinality of edge set $E(G_{dev})$, which determines the density of $G_{dev}$, is inversely analogous to cardinality of $V(G_{map})$.

\newpage

\begin{definition}{}
For a given set of mobile devices, say, $dev$, that are moving inside a city represented by its corresponding $G_{map}$ we define the density of this network (i.e., $G_{dev}$), denoting it with $D$ as follows:
\begin{equation}\label{density}
    D(dev, G_{map})=\dfrac{|I|+|S|}{|V(G_{map})|},
\end{equation}

\noindent where $|I|$ and $|S|$ correspond to the number of infected and susceptible devices respectively, and from which it follows that for a given number of devices, say, $n$, the higher the cardinality of $|V(G_{map})|$ the less the density of $G_{dev}$.

\end{definition}{}

\vspace*{0.1in}
\section {Evaluation}
\label{sec:Evaluation}
\vspace*{0.05in}

\noindent In this section, we present the setup of our experiments performed by our simulator that deploys our malware propagation and device mobility models. We perform a series of simulations focusing mainly on the effect of response-time demanded by a counter-measure to be activated in order to clean up the infected devices, minimizing or eliminating the {\it IS-rate}, i.e., $R(I,S)=0$.

\vspace*{0.1in}
\subsection{Experimental Design}
In the framework of this paper, we are interested in investigating the effect of the time a counter-measure needs to be activated (i.e., response-time) on the malware's propagation when it is been triggered by an after-infection time limit and also, in a second level, how other factors such as the size of a malware, the density of the network and the initial infected population affect the spread of malware. Specifically, we distinguish two categories of experiments both investigating the effect of the counter-measure's response-time on the spread of the malware: the first category concerns the activation of the counter-measure on each device by setting the response-time to various intervals, while the second one concerns the effect of other factors, such as the density of the network, and the initial size of the infected population.

In our experiments, we utilize the image of a city taken from Google Maps, transforming it to a black and white matrix, as we described in the previous section. Within this approach, we assign weights (attraction levels) to each cell $(i,j)$ with value $0$; recall that, cells with such values represent points on a road. Based on these values, we compute shortest paths for a set of start-destination points for each mobile device utilizing a shortest path algorithm. Next we present results for a series of experiments for malware consisted by $3$ and $6$ packets for various response time intervals and {\it IS-rate} $R(I,S)=0.25$ having an initially infected population consisted by $20$ devices and and initial susceptible population consisted by $80$ devices, i.e., $I=20$ and $S=80$.

\begin{figure}[t!]
\hrule\medskip\medskip
  \begin{minipage}[b]{0.5\linewidth}
    \centering
    \includegraphics[width=.95\linewidth]{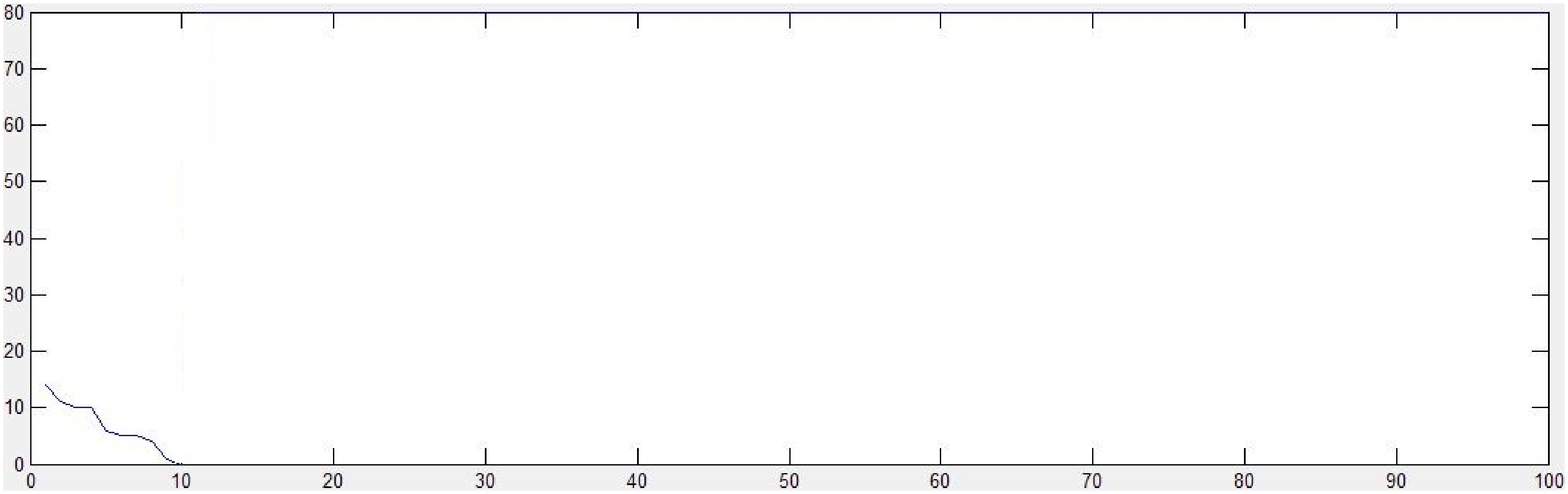}\\
     {\small $(a)$ $p=3, R_t \in [1, 5], R(I,S)=0.25$}
    \vspace{1ex}
  \end{minipage}
  \begin{minipage}[b]{0.5\linewidth}
    \centering
    \includegraphics[width=.95\linewidth]{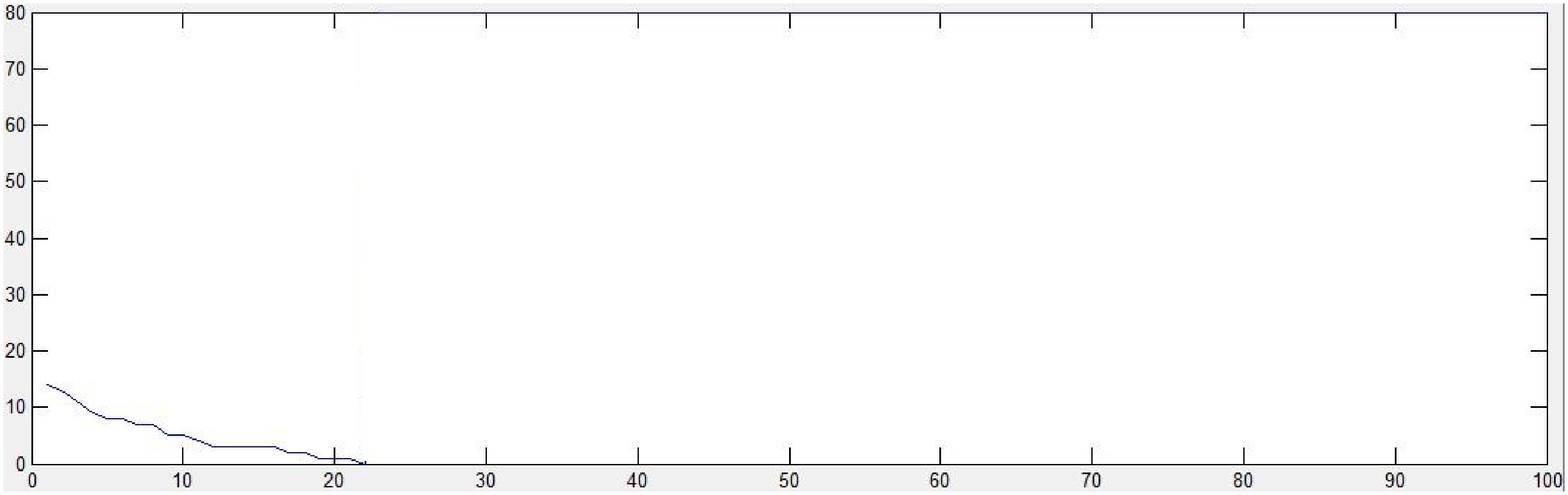}\\
     {\small $(b)$ $p=6, R_t \in [1, 5], R(I,S)=0.25$}
    \vspace{1ex}
  \end{minipage}

 \begin{minipage}[b]{0.5\linewidth}
    \centering
    \includegraphics[width=.95\linewidth]{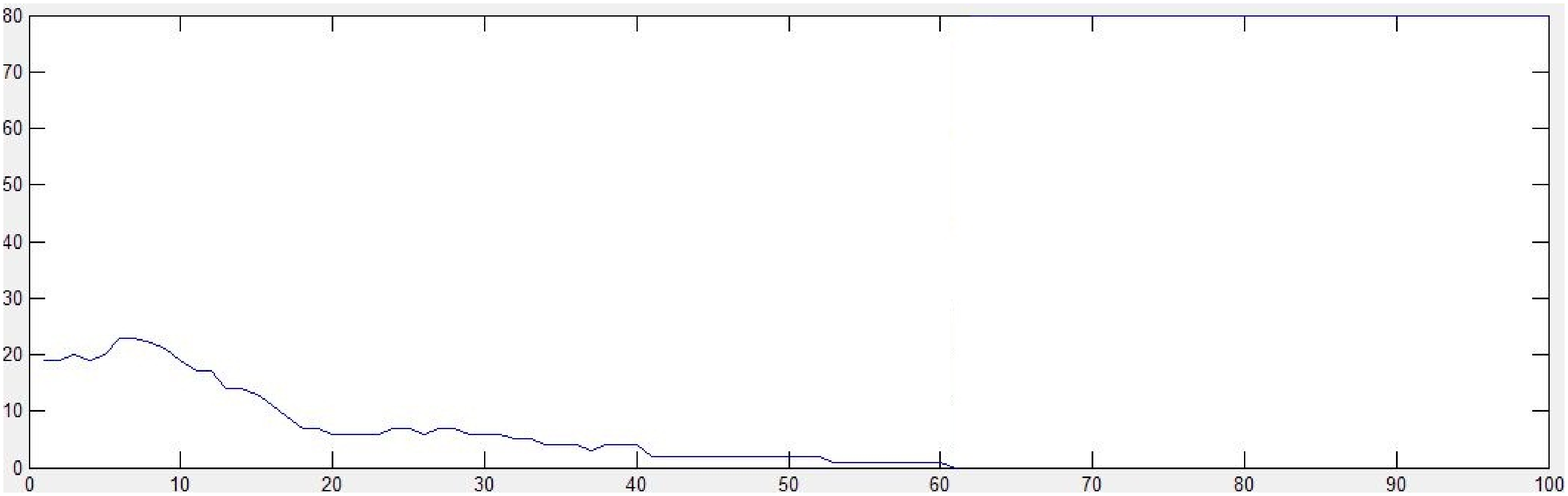}\\
     {\small $(c)$ $p=3, R_t \in [6, 10], R(I,S)=0.25$}
    \vspace{1ex}
  \end{minipage}
  \begin{minipage}[b]{0.5\linewidth}
    \centering
    \includegraphics[width=.95\linewidth]{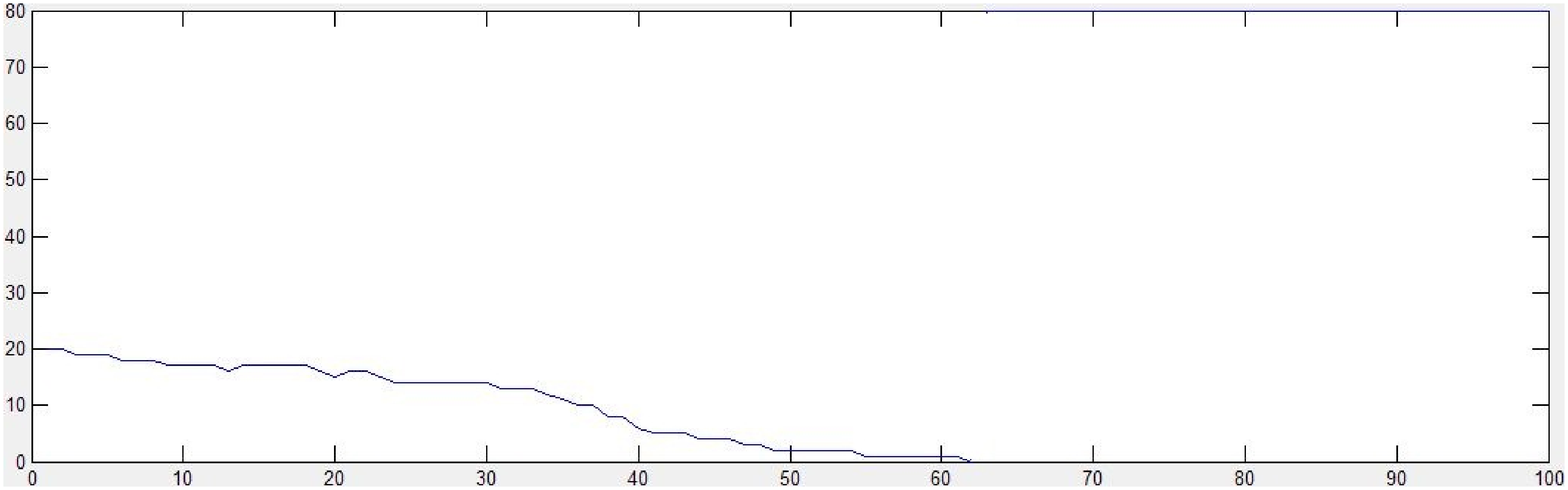}\\
     {\small $(d)$ $p=6, R_t \in [6, 10], R(I,S)=0.25$}
    \vspace{1ex}
  \end{minipage}

   \begin{minipage}[b]{0.5\linewidth}
    \centering
    \includegraphics[width=.95\linewidth]{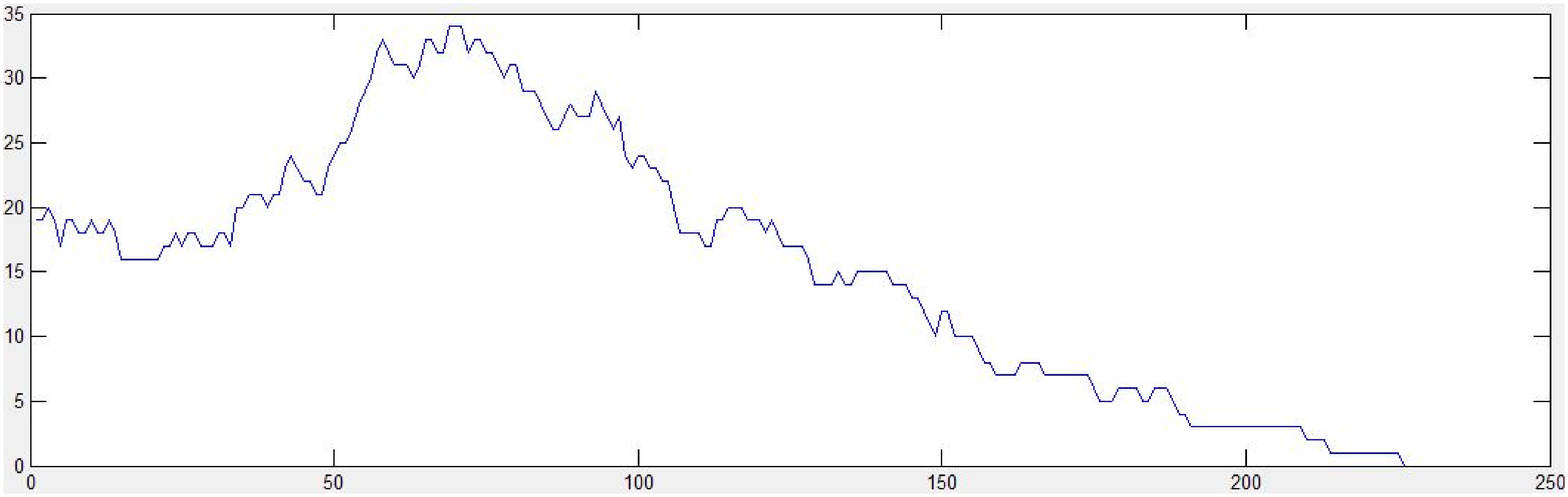}\\
     {\small $(e)$ $p=3, R_t \in [11, 20], R(I,S)=0.25$}
    \vspace{1ex}
  \end{minipage}
  \begin{minipage}[b]{0.5\linewidth}
    \centering
    \includegraphics[width=.95\linewidth]{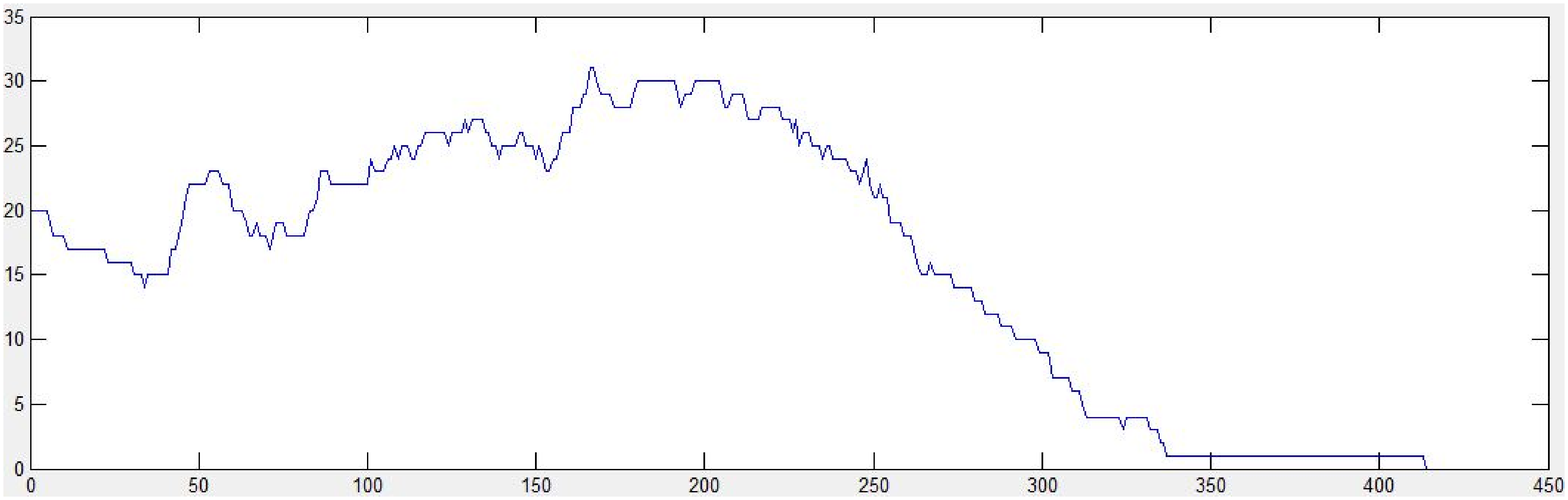}\\
     {\small $(f)$ $p=6, R_t \in [11, 20], R(I,S)=0.25$}
    \vspace{1ex}
  \end{minipage}

 \begin{minipage}[b]{0.5\linewidth}
    \centering
    \includegraphics[width=.95\linewidth]{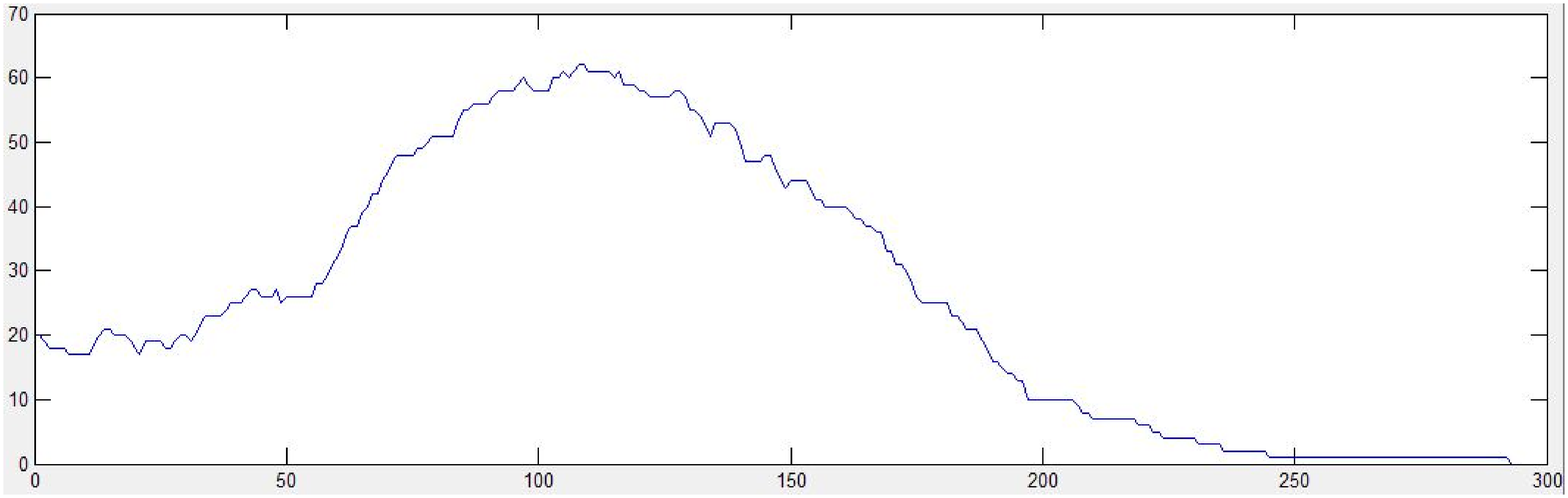}\\
     {\small $(g)$ $p=3, R_t \in [21, 40], R(I,S)=0.25$}
    \vspace{1ex}
  \end{minipage}
  \begin{minipage}[b]{0.5\linewidth}
    \centering
    \includegraphics[width=.95\linewidth]{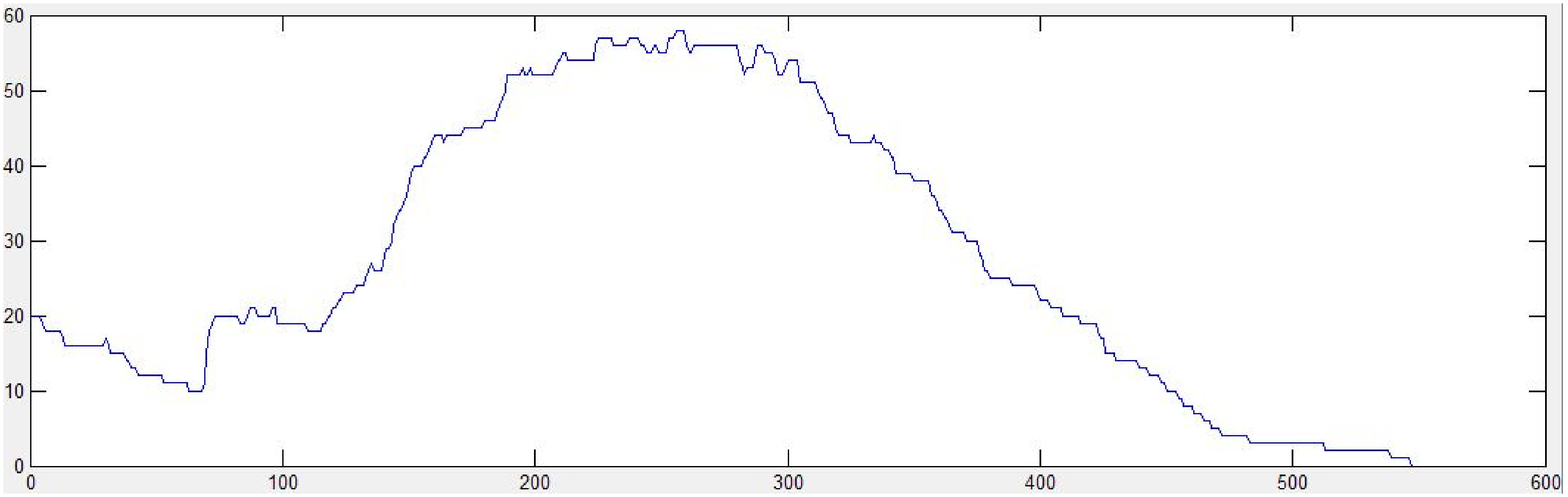}\\
     {\small $(h)$ $p=6, R_t \in [21, 40], R(I,S)=0.25$}
    \vspace{1ex}
  \end{minipage}

   \begin{minipage}[b]{0.5\linewidth}
    \centering
    \includegraphics[width=.95\linewidth]{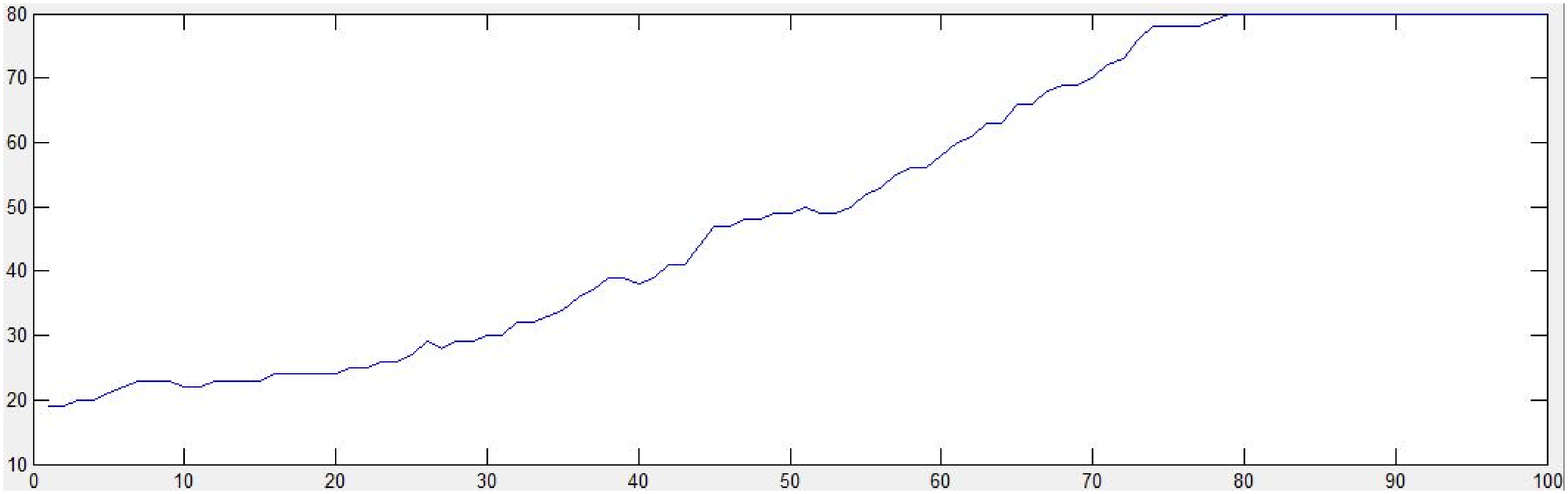}\\
     {\small $(i)$ $p=3, R_t \in [41, 80], R(I,S)=0.25$}
    \vspace{1ex}
  \end{minipage}
  \begin{minipage}[b]{0.5\linewidth}
    \centering
    \includegraphics[width=.95 \linewidth]{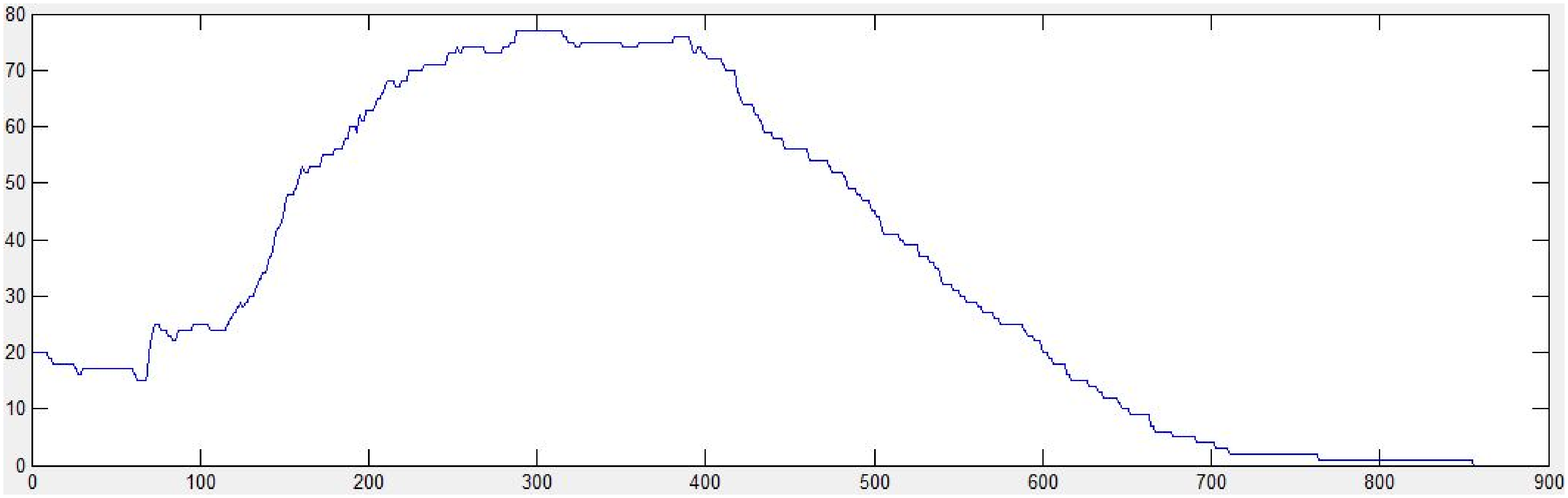}\\
     {\small $(j)$ $p=6, R_t \in [41, 80], R(I,S)=0.25$}
    \vspace{1ex}
  \end{minipage}

  \medskip\hrule\smallskip
  \caption{\small{Simulation experiments for different values of malware packets ($p$) and various counter-measure response times ($R_t$).}}
  \label{fig:intervals}
\end{figure}

In the two categories of experiments presented, we study different intervals of values concerning the response-time of a counter-measure; note that, not all the devices have the same response-time but on each one is assigned a response-time of the interval under consideration. Additionally, using the same categories of experiments we investigate the effect of the malware size, expressed in packets, and show how it can affect the result on both categories. In both categories of experiments, we perform a set of simulations on malware spread between moving devices in a city region. In the first category we keep the same ratio of the initial infected population and susceptible devices varying the counter-measure response time, while in the second one we increase the initial infected population and the network's density.

Throughout the paper, we shall denote the initial infected population by $I$, the susceptible devices by $S$, the counter-measure response time by $R_t$ and the malware's size by $p$. To this point we ought to notice that the $R_t=t$ does not actually corresponds to $t$ simulation steps but in the case where the size $p$ of the malwere (in terms of packs) is greater that 1 it holds $R_t = t \times p$, that means, the infected device will propagate $t$ times the full malware, and thus $t \times p$ simulation steps are required before its counter-measure been activated.
In Figures~\ref{fig:intervals}, \ref{fig:infected} and \ref{fig:density}, the $x-axis$ refers to the simulation steps taken up to the end of simulation, while the $y-axis$ refers to the number of infected devices.

\vspace*{0.1in}
\subsection{Pandemic Prevention for Various Response-time Intervals}
As we can observe in Figures~\ref{fig:intervals}($a$) and \ref{fig:intervals}($b$), the size of the spreading malware does not affect the spread at all since the response time is low. That means, the propagation of malware to neighboring susceptible devices fails, despite its size, due to the early activation of the counter-measure. However, the duplication of the size of the malware leads to the duplication on the time required for all the susceptible devices in the city to avoid the infection and all infected ones get sanitized. Moreover, increasing the counter-measure response time inside the same order of magnitude, we observe that still there is no increase on the number of the infected population, where in both cases (see, Figures~\ref{fig:intervals}($c$) and \ref{fig:intervals}($d$)) the number of infected population, from the start of simulation, is monotonically decreasing since the cure has started once the counter measure has been activated.

On the other hand, leaving the rest parameters unchanged and increasing only the response-time of the counter-measure, we observe a global maximum of the spread in both experiments (see, Figures~\ref{fig:intervals}($e$) -- \ref{fig:intervals}($f$) and Figures~\ref{fig:intervals}($g$) -- \ref{fig:intervals}($h$)) which are achieved on the 70th and $160$th simulation steps, and on the $110$th and $260$th simulation steps respectively. In both cases the maxima are due to the multiplication of the response-time $R_t$ by a factor $f$ that causes the infected devices to propagate the full malware $f$ times more than in the previous experiments. However in both cases the malware failed to spread to all the population since for the specific parameters (i.e., $p=3$ and $p=6$) the counter-measure activated early enough and thus the pandemic is prevented.

Finally, in Figure~\ref{fig:intervals}($i$) we can see that further increase of the response-time interval, in the case of a malware of smaller size (i.e., $p=3$), lets the malware spread to all the susceptible devices. However, for the same response-time interval but with double malware size (i.e., $p=6$) (see, Figure~\ref{fig:intervals}($j$)), we observe that the infection failed to spread to all the susceptible devices. Note that, an increase on the response-time facilitates the propagation of small size malware.

\vspace*{0.1in}
\subsection{Other Results on Pandemic Prevention}
Next we present results for a series of experiments for malware consisted by $3$ and $6$ packets for various response time intervals. We change some factors concerning the characteristics of the network (i.e., $G_{dev}$) and firstly perform a series of experiments for a different {\it IS-rate} $R(I,S)=0.66$ having an initially infected population consisted by $40$ devices and and initial susceptible population consisted by $60$ devices, i.e., $I=40$ and $S=60$, while then we increase the density $D$ of our network by duplicating the number of devices keeping the {\it IS-rate} equal to that of Figure~\ref{fig:intervals}; recall that, $R(I,S)=0.25$, where $I=40$ and $S=160$.

\begin{figure}[t!]
\hrule\medskip\medskip
   \begin{minipage}[b]{0.5\linewidth}
    \centering
    \includegraphics[width=.95\linewidth]{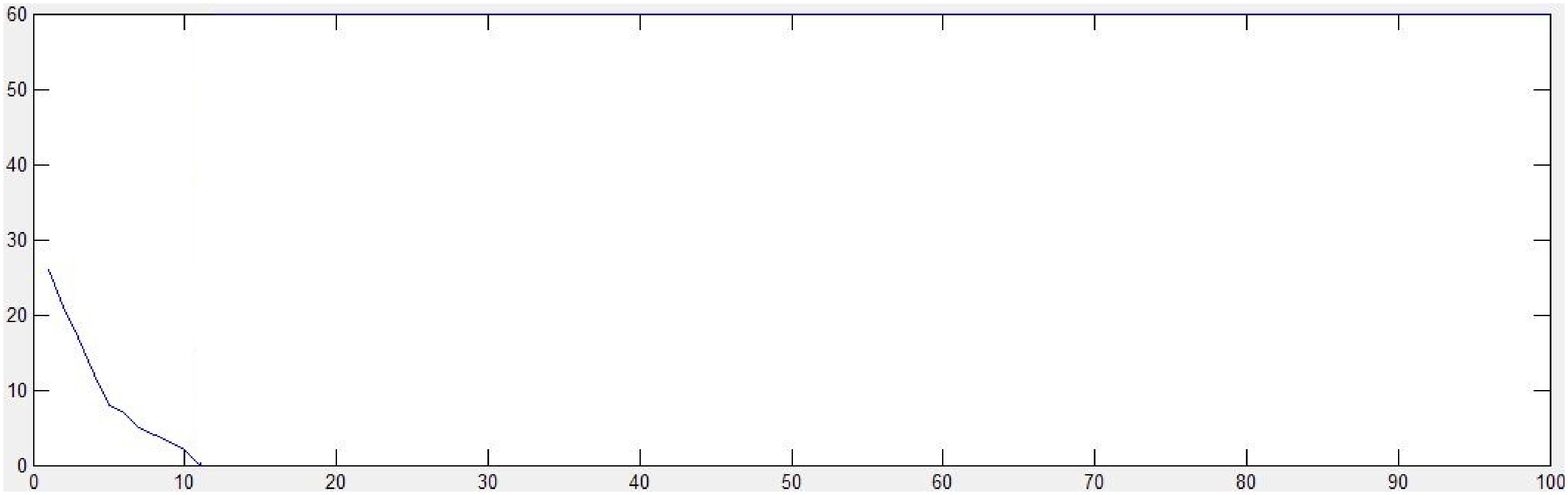}\\
     {\small $(a)$ $p=3, R_t \in [1, 5], R(I,S)=0.66$}
    \vspace{1ex}
  \end{minipage}
  \begin{minipage}[b]{0.5\linewidth}
    \centering
    \includegraphics[width=.95 \linewidth]{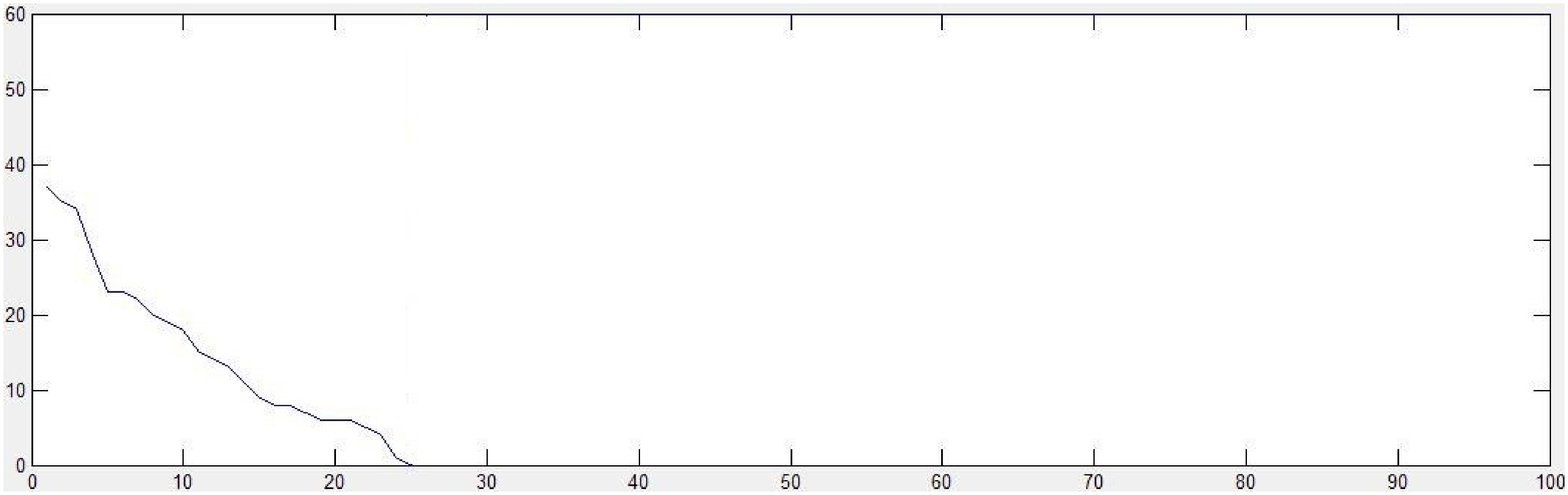}\\
     {\small $(b)$ $p=6, R_t \in [1, 5], R(I,S)=0.66$}
    \vspace{1ex}
  \end{minipage}

\medskip\smallskip
  \begin{minipage}[b]{0.5\linewidth}
    \centering
    \includegraphics[width=.95\linewidth]{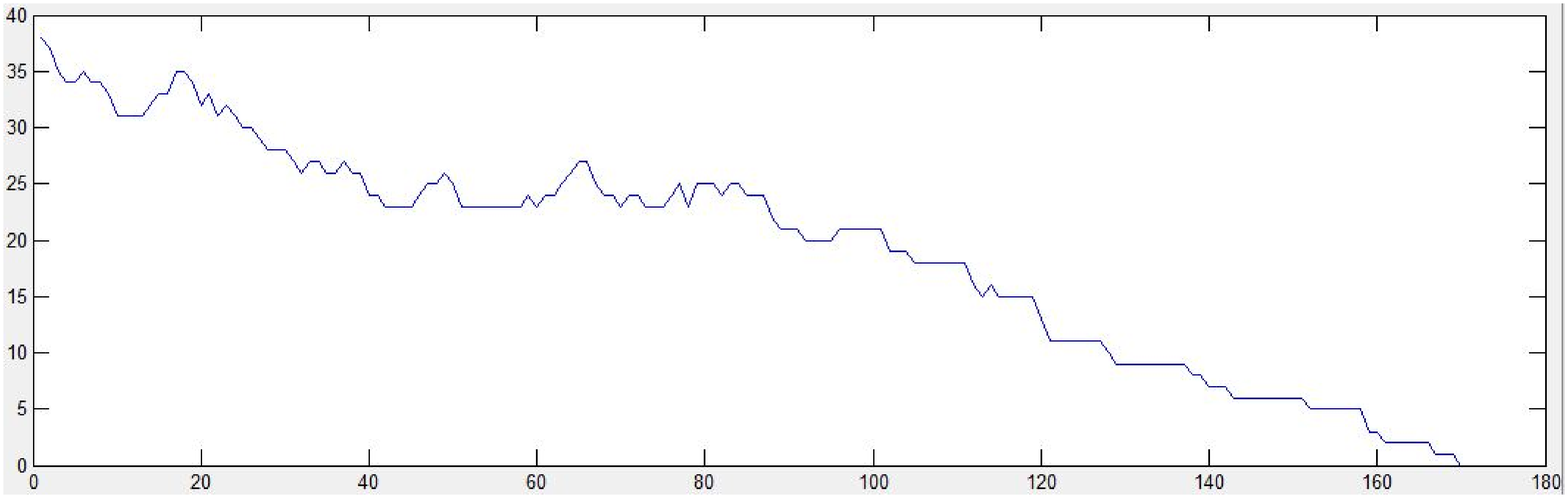}\\
     {\small $(c)$ $p=3, R_t \in [11, 20], R(I,S)=0.66$}
    \vspace{1ex}
  \end{minipage}
  \begin{minipage}[b]{0.5\linewidth}
    \centering
    \includegraphics[width=.95\linewidth]{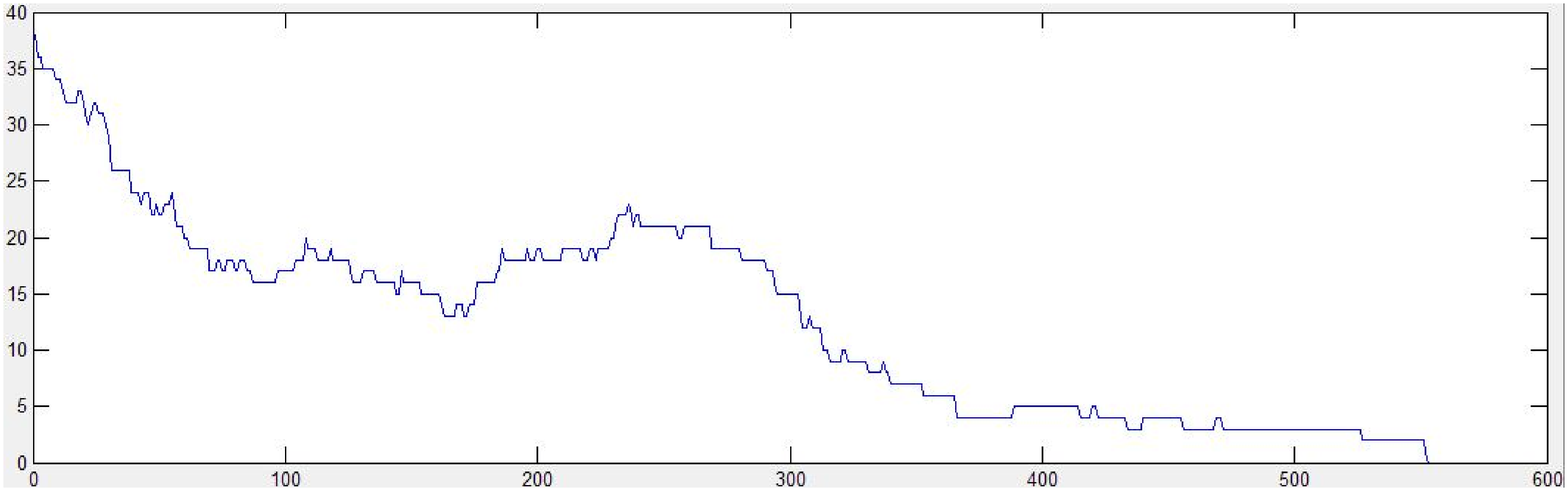}\\
     {\small $(d)$ $p=6, R_t \in [11, 20], R(I,S)=0.66$}
    \vspace{1ex}
  \end{minipage}

\medskip\smallskip
  \begin{minipage}[b]{0.5\linewidth}
    \centering
    \includegraphics[width=.95\linewidth]{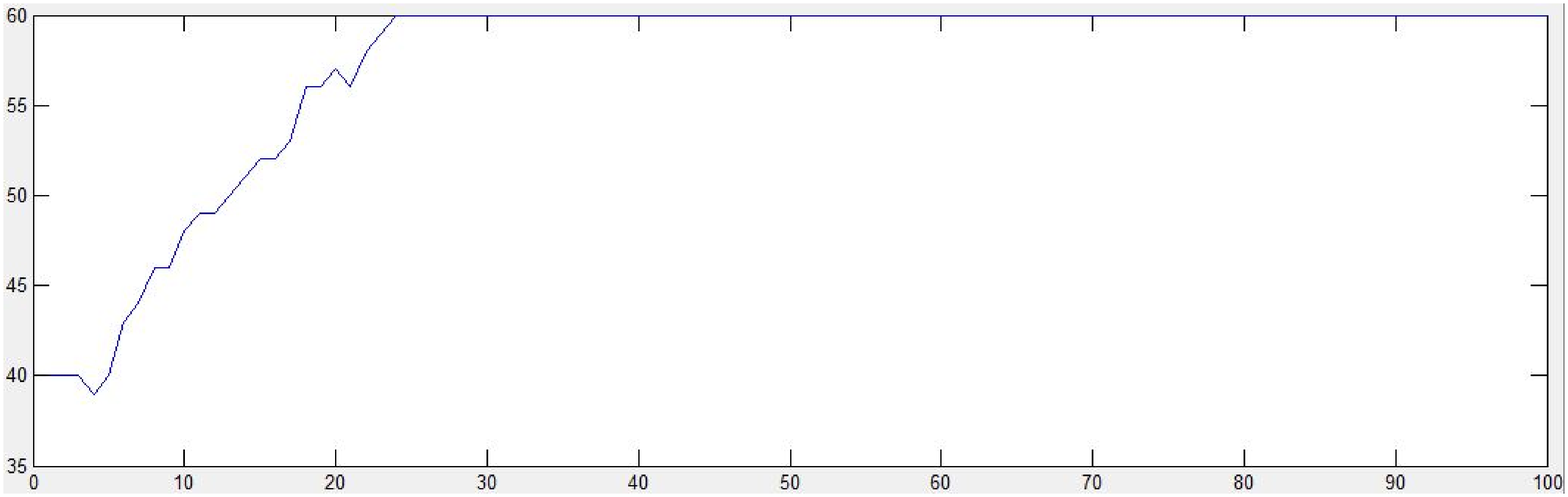}\\
     {\small $(e)$ $p=3, R_t \in [41, 80], R(I,S)=0.66$}
    \vspace{1ex}
  \end{minipage}
  \begin{minipage}[b]{0.5\linewidth}
    \centering
    \includegraphics[width=.95\linewidth]{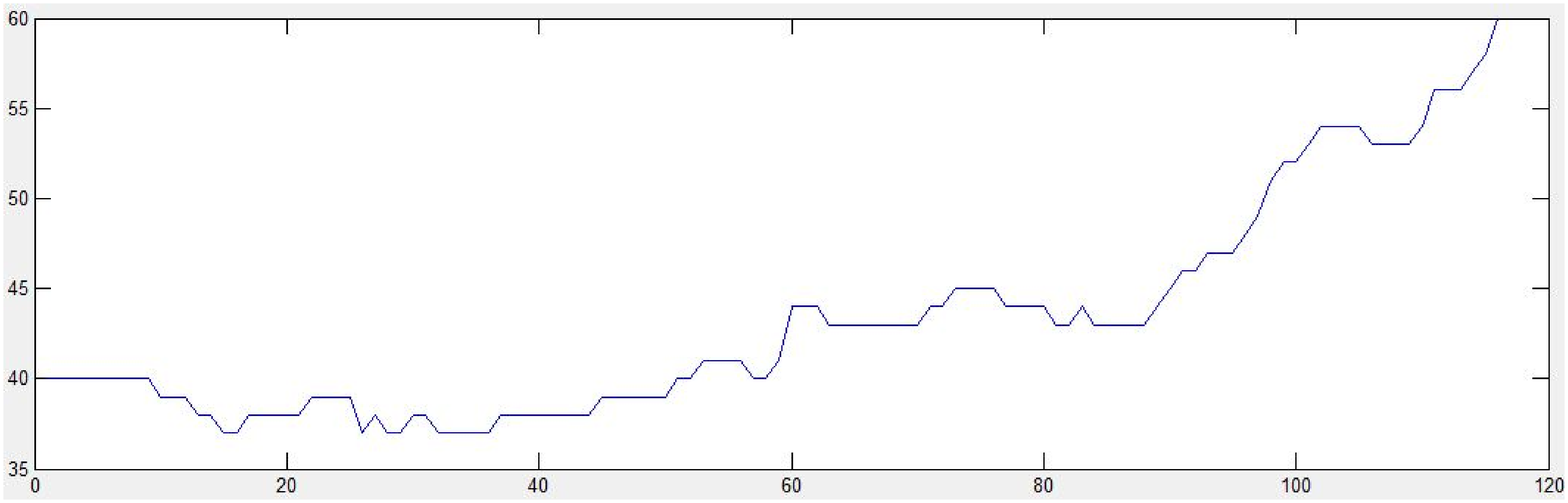}\\
     {\small $(f)$ $p=6, R_t \in [41, 80], R(I,S)=0.66$}
    \vspace{1ex}
  \end{minipage}

  \medskip\hrule\smallskip
  \caption{\small{Simulation experiments for the cases of a malware with $p=3$ and $p=6$ for  a network with double initially infected devices.}}
  \label{fig:infected}
\end{figure}

\begin{figure}[t!]
\hrule\medskip\medskip
   \begin{minipage}[b]{0.5\linewidth}
    \centering
    \includegraphics[width=.95\linewidth]{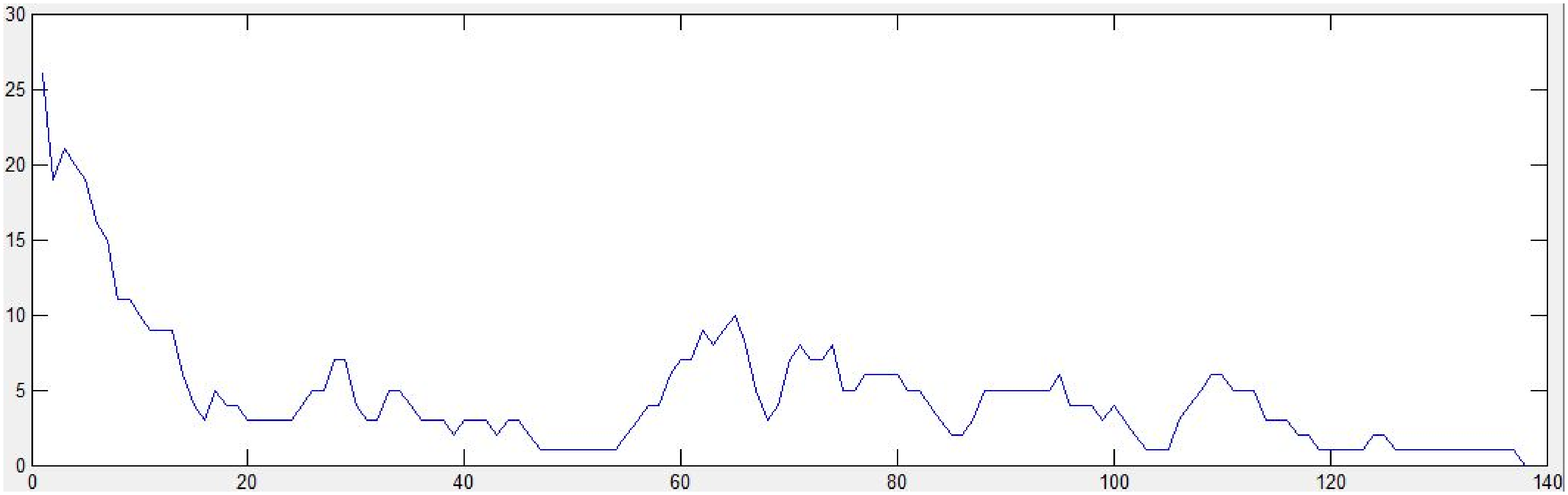}\\
     {\small $(a)$ $p=3, R_t \in [1, 5], R(I,S)=0.25$}
    \vspace{1ex}
  \end{minipage}
  \begin{minipage}[b]{0.5\linewidth}
    \centering
    \includegraphics[width=.95 \linewidth]{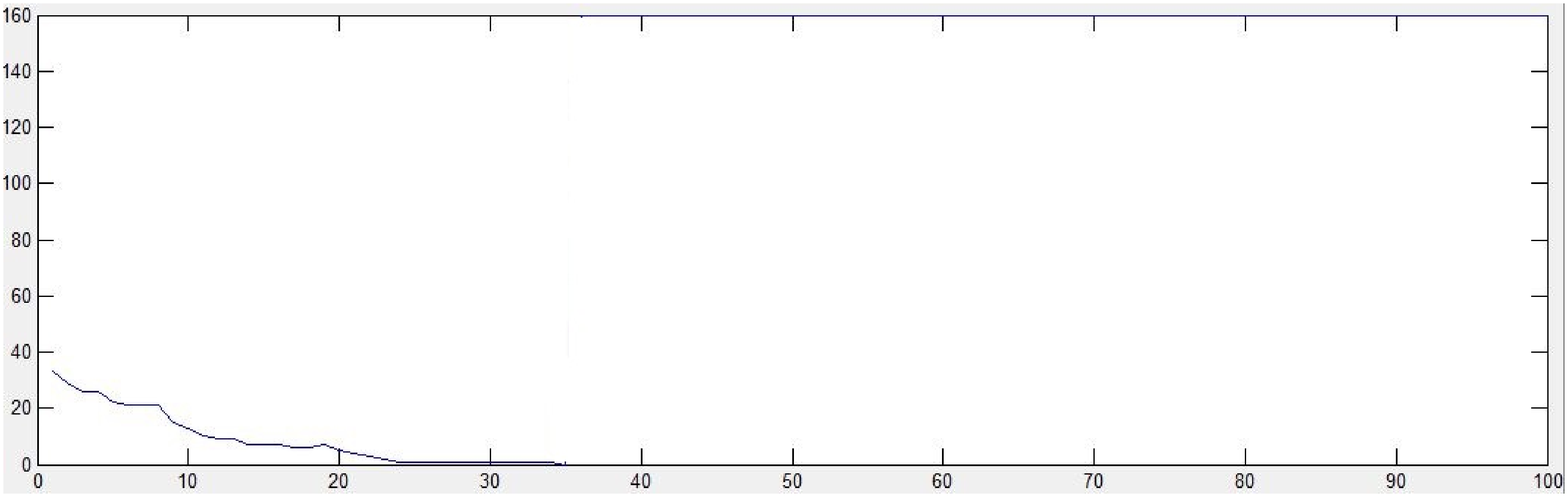}\\
     {\small $(b)$ $p=6, R_t \in [1, 5], R(I,S)=0.25$}
    \vspace{1ex}
  \end{minipage}

\medskip\smallskip
  \begin{minipage}[b]{0.5\linewidth}
    \centering
    \includegraphics[width=.95\linewidth]{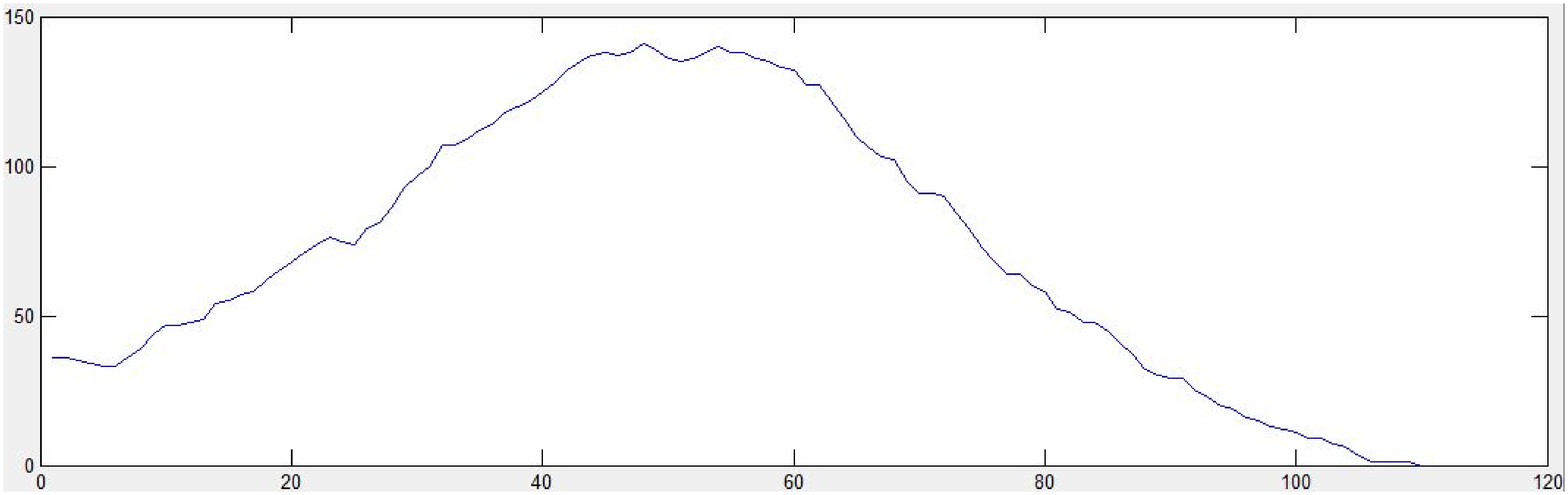}\\
     {\small $(c)$ $p=3, R_t \in [11, 20], R(I,S)=0.25$}
    \vspace{1ex}
  \end{minipage}
  \begin{minipage}[b]{0.5\linewidth}
    \centering
    \includegraphics[width=.95\linewidth]{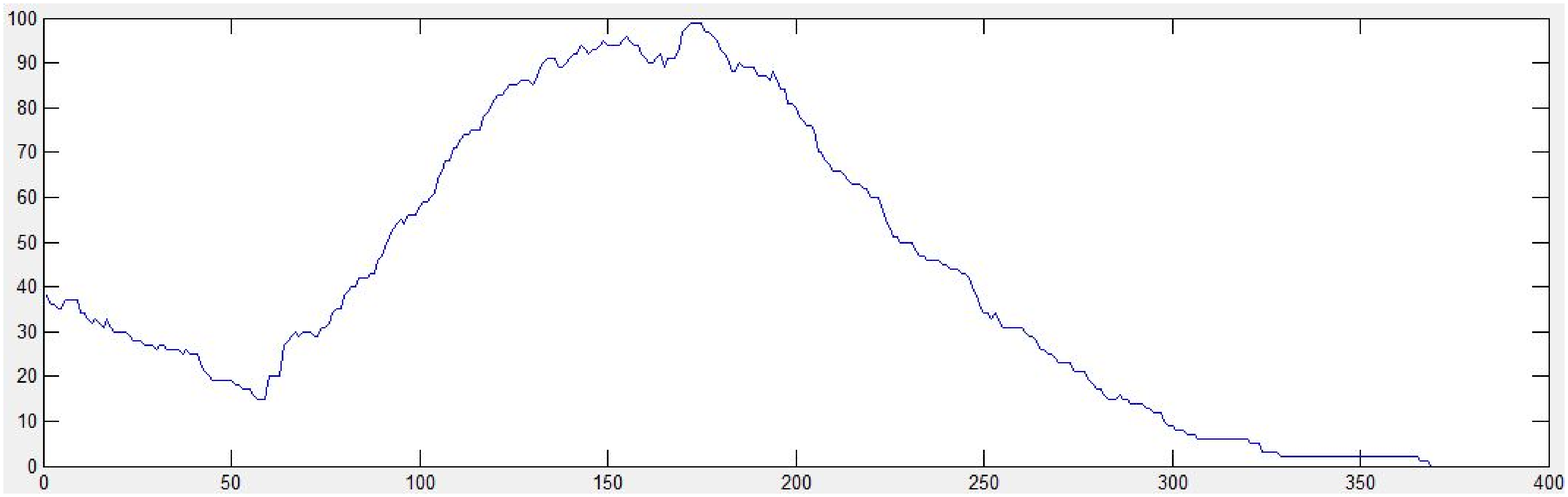}\\
     {\small $(d)$ $p=6, R_t \in [11, 20], R(I,S)=0.25$}
    \vspace{1ex}
  \end{minipage}

\medskip\smallskip
  \begin{minipage}[b]{0.5\linewidth}
    \centering
    \includegraphics[width=.95\linewidth]{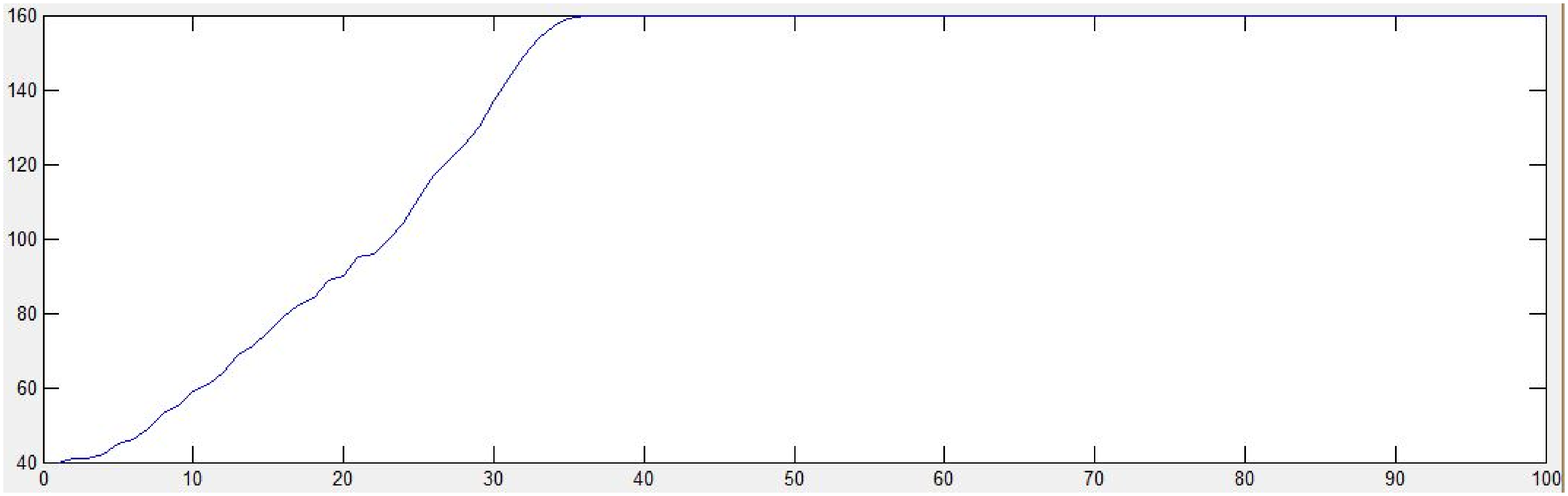}\\
     {\small $(e)$ $p=3, R_t \in [41, 80], R(I,S)=0.25$}
    \vspace{1ex}
  \end{minipage}
  \begin{minipage}[b]{0.5\linewidth}
    \centering
    \includegraphics[width=.95\linewidth]{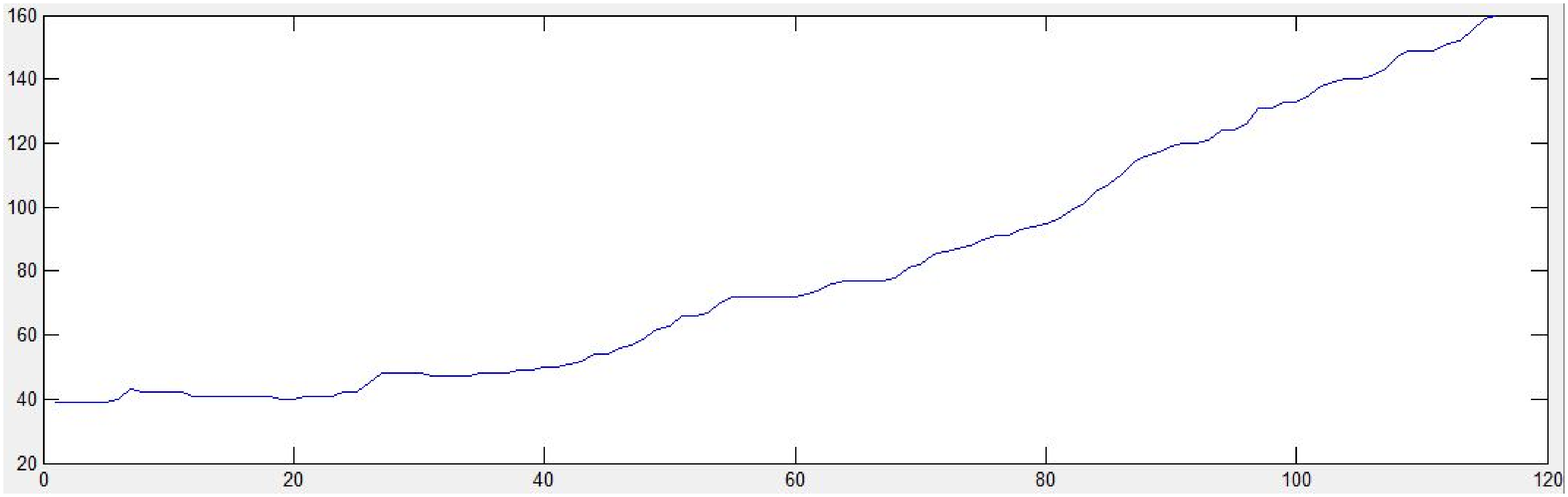}\\
     {\small $(f)$ $p=6, R_t \in [41, 80], R(I,S)=0.25$}
    \vspace{1ex}
  \end{minipage}

  \medskip\hrule\smallskip
  \caption{\small{Simulation experiments for the cases of a malware with $p=3$ and $p=6$  a network  with double density.}}
  \label{fig:density}
\end{figure}

The results of the second category of our experiments are depicted in Figures~\ref{fig:infected} and \ref{fig:density}. In this category, we modify the experiments of Figures~\ref{fig:intervals}($a$) -- \ref{fig:intervals}($b$), \ref{fig:intervals}($e$) -- \ref{fig:intervals}($f$) and \ref{fig:intervals}($i$) -- \ref{fig:intervals}($j$), changing the ratio $I/S$ from $0.25$ to $0.66$ (see, Figure~\ref{fig:infected}) and the density $D$ of the network by duplicating the number of devices (see, Figure~\ref{fig:density}); recall that, $I$ and $S$ denote the number of the initial infected and susceptible population, respectively.

It is rational to expect that in presence of an early activated counter-measure (i.e., response-time intervals close to $0$) a quick response is crucial for the immediate repression of malware's propagation. So, response-time intervals close to $0$, act the same for pandemic prevention despite the size of the initial infected population, as shown in the results contrasting Figures~\ref{fig:infected}($a$) and \ref{fig:infected}($b$) with Figures~\ref{fig:intervals}($a$) and \ref{fig:intervals}($b$), where the prevention of a pandemic needs almost the same number of simulation steps for these couples of experiments (see, Figures~\ref{fig:infected}($a$) -- \ref{fig:intervals}($a$) and \ref{fig:infected}($b$) -- \ref{fig:intervals}($b$)).

Comparing Figures~\ref{fig:infected}($c$) and \ref{fig:infected}($d$) with Figures~\ref{fig:intervals}($e$) and \ref{fig:intervals}($f$), respectively, we observe that in the first case the flow of the propagation follows a decreasing monotonicity, with less or none grows (see, Figure~\ref{fig:infected}($c$) and Figure~\ref{fig:intervals}($e$) for a $3$-packet malware), that is attributed  to the number of available susceptible devices as also to the activation of the counter-measure that involves the immunization of the infected host. Additionally, comparing Figure~\ref{fig:infected}($d$) with Figure~\ref{fig:intervals}($f$), where the size of the spreading malware is duplicated, we observe a slower growth on the number of infected devices (grow-level). This observation can be explained by the fact that as more packets needed to infect a device it slows down the infection and, thus, the activation of the the counter-measure requires more simulation steps. So, the infected devices with response-time near the upper bound of the assigned interval have more time to infect other susceptible devices and, hence, in this case the whole pandemic prevention of Figure~\ref{fig:infected}($d$) demands more simulation steps than in the case depicted in Figure~\ref{fig:intervals}($f$) (i.e., $550$ and $430$, respectively).

However, comparing Figure~\ref{fig:infected}($e$) and \ref{fig:infected}($f$) with Figure~\ref{fig:intervals}($i$) and \ref{fig:intervals}($j$), respectively, we observe interesting evidences about our intuition that the number of initially infected devices could significantly speed up the propagation of malware in such a network. So, in the experiments depicted in Figure~\ref{fig:infected}($e$) and \ref{fig:infected}($f$) we observe that in both cases the counter-measure's activation, due to its larger response time, failed to prevent pandemic, in contrast to the case of experiments presented in Figure~\ref{fig:intervals}($i$) and \ref{fig:intervals}($j$), where while in Figure~\ref{fig:intervals}($i$) a malware of smaller size achieved to propagate infecting all the susceptible devices in the network, in Figure~\ref{fig:intervals}($j$) we observe that in contrast with Figure~\ref{fig:infected}($f$), the pandemic failed obviously due to the number of initial infected population.

In the second set of experiments for malware propagation in a network of mobile devices with double density of the network on which the experiments of Figure~\ref{fig:intervals} performed we observe a differentiation on the malware's spread behavior where, while in the case of a $6$-packet malware (see, Figures~\ref{fig:density}($b$) and \ref{fig:intervals}($b$)) no significant variations arise, in the case of a $3$-packet malware's spread (see, Figures~\ref{fig:density}($a$) and \ref{fig:intervals}($a$)), it is obvious that in the first case the increased number of initially infected devices acts subsidiary to the malware's spread making the propagation lasting longer as for a greater period of time (in terms of simulation steps) there still exist infected devices in the network.

Contrasting Figures~\ref{fig:intervals}($e$) and \ref{fig:intervals}($f$) with Figures~\ref{fig:density}($c$) and \ref{fig:density}($d$), we observe the same behavior in malware's spread, where in both cases the propagation exhibits a global maximum, however, a more detailed view, could reveal that in the first case (see, Figures~\ref{fig:density}($c$) and \ref{fig:density}($d$)), where the density of the network is duplicated the whole pandemic prevention lasts less than in the first one (see, Figures~\ref{fig:intervals}($e$) and \ref{fig:intervals}($f$)) in terms of simulation steps. Comparing Figure~\ref{fig:intervals}($e$) and \ref{fig:intervals}($f$) with Figure~\ref{fig:density}($c$) and \ref{fig:density}($d$), we observe that the network's density (as also the initial infected population, see Figures~\ref{fig:infected}($c$) and \ref{fig:infected}($d$)) do not affect the prevention of a pandemic when a properly activated counter-measure exists. This result is explained by the fact that since the counter-measure acts the same way in both cases, in the $50$th simulation step $20$ and $130$ devices have been sanitized (see, Figures~\ref{fig:intervals}($e$) and \ref{fig:density}($c$), respectively).

Finally, comparing Figures~\ref{fig:density}($e$) and \ref{fig:density}($f$) with Figures~\ref{fig:intervals}($i$) and \ref{fig:intervals}($j$), respectively, also interesting evidences arise about our intuition that an increase on the density of the network could significantly speed up the propagation of malware. More precisely, in Figures~\ref{fig:density}($e$) and \ref{fig:density}($f$) we observe that in both cases the counter-measure's activation, due to its larger response time, failed to prevent pandemic, in contrast to the case of experiments presented in Figures~\ref{fig:intervals}($i$) and \ref{fig:intervals}($j$), where while in Figure~\ref{fig:intervals}($i$) a malware of smaller size achieved to propagate infecting all the susceptible devices in the network, in Figure~\ref{fig:intervals}($j$), the pandemic failed obviously due to the malware's size.

\vspace*{0.1in}
\section {Concluding Remarks}
\label{sec:model_design}
\vspace*{0.05in}

\noindent In this work we investigate the effect of the response-time of a counter-measure and establish time bounds that prevent pandemic of a malware in mobile devices moving in an area of a city. We first proposed a malware propagation model to simulate the spread of malware based on geological proximity and, then, a device mobility model that generates traces for the mobile devices, utilizing shortest path algorithms, inside the city represented by its image taken from Google Maps.  Finally, we developed a simulator for malware spread among mobile devices, implementing the models proposed above, and performed a series of experiments providing results on the effect of counter-measure response-time on malware's spread.

In light of our proposed models in this paper it would be very interesting to explore the dynamics of other epidemic models such as the SIRp where an infected node can move to a repaired state (but still vulnerable). An interesting direction could be the investigation of cases where the counter-measure does not immunize the infected host against malware and hence the system occasionally may come under an endless stable equipoise between the infected and susceptible population.


\end{document}